# A Machine-Learning Accelerated Grand Canonical Sampling Framework for Nuclear Quantum Effects in Constant Potential Electrochemistry

Menglin Sun[1]†, Bin Jin[1]†, Xiaolong Yang[1]†, Shenzhen Xu[1,2]*

[1]School of Materials Science and Engineering, Peking University, Beijing 100871, People's Republic of China

[2]AI for Science Institute, Beijing 100084, People's Republic of China

† The authors contribute equally

* Corresponding author: xushenzhen@pku.edu.cn

## TOC Graphic:

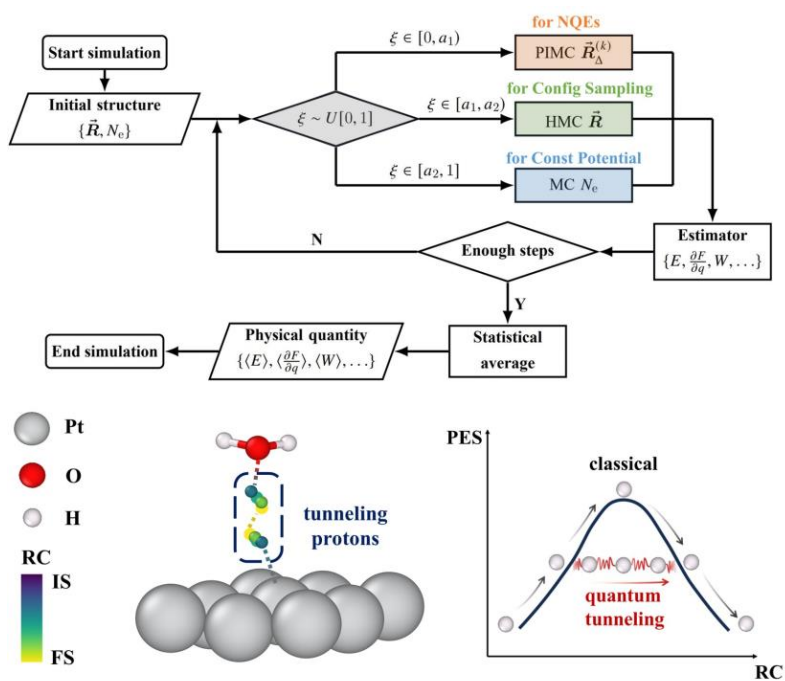

## Abstract

Proton-coupled electron transfer (PCET) is the key step for energy conversion in electrocatalysis. Atomic-scale simulation acts as an indispensable tool to provide a microscopic understanding of PCET. However, consideration of the quantum nature of transferring protons under an exact grand canonical (GC) constant potential condition is a great challenge for theoretical electrocatalysis. Here, we develop a unified computational framework to explicitly treat nuclear quantum effects (NQEs) by a sufficient GC sampling, further assisted by a machine learning force field adapted for electrochemical conditions. Our work demonstrates a non-negligible impact of NQEs on PCET simulations for hydrogen evolution reaction at room temperature, and provides a physical picture that wave-like quantum characteristic of the transferring protons facilitates the particles to tunnel through classical barriers in PCET paths, leading to a remarkable activation energy reduction compared to classical simulations. Moreover, the physical insight of NQEs may reshape our fundamental understanding of other types of PCET reactions in broader scenarios of energy conversion processes.

## Introduction

Understanding and controlling energy conversion processes are the key to the development of innovative green energy technologies. Electrochemical catalysis on various electrode materials facilitates efficient conversion between transient electrical energies and stable chemical energies, for example, electrocatalytic hydrogen evolution reactions (HER)[1] during water splitting, oxygen reduction reactions (ORR)[2] in fuel cells, electrochemical $CO_2$ reduction reactions ($CO_2$RR)[3] which simultaneously mitigates $CO_2$ emission and produces fuels. Proton-coupled electron transfer (PCET), an essential elementary reaction step shared by most of the electrochemical energy conversion systems, determines the reaction rates, efficiencies, and selectivity of different electrocatalytic cells. Deep insights and understanding of the proton transfer step in complex environment near the electrode/solvent interfaces thus draw great attention and interests from the community of electrochemistry and chemical physics.

For a typical PCET process in electrochemical systems of metal/aqueous solution interfaces, a proton originally from $H_2O$ network transfers toward catalytic metal sites or a certain reactant species, accompanied by electron transfer (via electrode surfaces) under a constant potential condition. Detecting and observing these intricate and localized events at a microscopic level presents a challenge for experimental methods, across both spatial and temporal dimensions. However, atomic simulations offer a powerful alternative, allowing us to examine the kinetics and thermodynamics of elementary PCET steps. From the above conceptual description of PCET, we can obtain three key requirements that would be critical in theoretical modeling: (1) a sufficient sampling of complex electrode/solvent interfacial environment; (2) a constant potential condition corresponding to a grand canonical (GC) ensemble; (3) a physically appropriate and exact treatment for protons involved in a PCET process.

Electrochemical theorists have developed many different approaches for modeling PCET reactions in the past decades. One of the most widely used methods is the computational hydrogen electrode (CHE) model[4] originally proposed by Nørskov and colleagues, by which the thermodynamic reaction energies under a certain applied potential can be computed referred to the standard hydrogen electrode (SHE) or the reversible hydrogen electrode (RHE). Extension to kinetic barrier calculations considering a constant potential condition was achieved in 2016.[5,6] The charge extrapolation method adjusts potential energy surfaces (PES) that are initially calculated with a constant charge model. This adjustment accounts for the electric double layer (EDL) at electrode/solvent interfaces, using a capacitor model to approximate the *a posteriori* correction terms. The great advantages of low computational cost and easy implementation make the above approaches popular in the electrochemistry community of PCET simulations. However, it is widely acknowledged that there are shortcomings in the approach, which considers only the initial states (IS), final states (FS), and transition states (TS) along a specific single path of the PCET steps, leading to lack of configurational sampling. The charge extrapolation method, although applying a constant potential correction term, is not an exact treatment for the constant potential GC ensemble, one issue is that the TS is actually optimized under a constant charge model.

Regarding the necessity of sufficient configurational sampling involving electrode surface, reactant species and solvent (usually $H_2O$) environment, several groups employed first-principles constrained molecular dynamics (CMD)[7,8] to calculate mean forces (or free energy gradients) at different reaction coordinates (RC) along a reaction path,[9–11] in which the RC is predefined to describe reactions' progress. Configurational sampling is indeed performed by this method, while the expensive *ab initio* calculations limit the total MD steps along a trajectory (only ~ $10^4$ steps), the sufficiency of statistical sampling thus is still questionable. This limitation actually could be addressed by employment of machine learning potentials (MLPs), which will be presented and discussed in this work. CMD itself cannot provide a constant potential GC ensemble condition, two strategies were employed by previous studies, one is adding *a posteriori* constant potential correction term derived from the charge extrapolation scheme,[12] the other strategy is using the so-called grand canonical density functional theory (GC-DFT) method,[13,14] in which the Fermi level referred to the vacuum level (or the work function) is fixed at a certain value. This can be accomplished by iteratively varying the system's total number of electrons in self-consistent field (SCF) loops until the system's work function is converged to the targeted value. However, fixing the work function of each sampled configuration actually does not resemble the microstates' distribution corresponding to an exact GC ensemble, which will be discussed later in this work. Moreover, this GC-DFT approach could induce an issue when dealing with quantized protons under a GC ensemble when the system's total electron number is allowed to vary, please refer to **Supporting Information (SI) Section 1** for detailed discussion.

Proton is the key transferring species in PCET reactions. Since hydrogen is the lightest element in nature, it is well known that the nuclear quantum effects (NQEs) dominate the mechanism of proton transfers at extremely low temperatures. Most of previous theoretical work on PCET steps in electrochemical catalysis did not consider NQEs, partially due to the relatively high temperature of a standard condition (~ 300 K) and the expensive computational cost required. However, bunch of earlier studies have revealed that the quantum nature of protons plays an important role at room temperature in proton transfer processes in liquid water,[15] small organic molecules,[16,17] biological macro molecules,[18] and even crystalline oxide materials.[19] It is therefore inspiring to raise a question about the impact of NQEs in PCET steps of electrocatalysis. Hammes-Schiffer group pioneered the investigation on the quantum effect of homogeneous and heterogeneous PCET.[20] The computational method developed by the group provides a Fermi golden rule expression[21] of PCET reaction rates, in which vibronic states (the direct product of electronic states and proton vibrational states) are employed to compute the vibronic coupling term.[22] Both of the nonadiabatic effect and NQEs can be incorporated under this theoretical framework. However, a sufficient configurational sampling of the electrochemical environment at the electrode/solvent interface is lacking. Moreover, the approximated quantum treatment of protons in PCET, in which 1-D wavefunctions are solved along the oxidized and reduced diabatic PES of proton transfers,[20] is not fully exact. As suggested by Hammes-Schiffer group[20] and well acknowledged by the quantum dynamics community, the Feynman path integral (PI) algorithm[23,24] is a more exact simulation approach for proton quantization, which will be employed in this work.

Our goal is to investigate the NQEs of PCET under a constant potential electrochemical condition, which can be conducted by a unified statistical sampling framework developed by this study. We employ a grand canonical hybrid Monte Carlo (GC-HMC) algorithm to equilibrate the system with an external electronic reservoir at a certain chemical potential of electrons, and combine with a path integral Monte Carlo (PIMC) method to take account of NQEs. We also develop a MLP adapted for electrochemical simulations on the basis of the Deep Potential (DP),[25] in which the total electron number of the interface system can be adjusted to realize a GC sampling, and we refer to this MLP as DP-$N_e$ throughout this paper. This DP-$N_e$ MLP significantly improves the computational efficiency while maintaining a DFT level accuracy, enabling sufficient statistical sampling in our work with an affordable computational cost.

Electrocatalytic HER is a notable green-energy technology producing clean fuels,[26] it consists of multiple possible PCET steps, which is an ideal modeling system in electrochemistry. We thus are interested in the NQEs on the thermodynamics and kinetics of elementary PCET steps of HER in this study. The Volmer step ($H^+_{sol}$ + $e^-$ + * → H*, where $H^+_{sol}$ represents a solvated proton in water solution, $e^-$ comes from a cathode, and "*" means a surface site at cathode surface) and the Heyrovsky step ($H^+_{sol}$ + $e^-$ + H* → $H_2$) are the two fundamental PCET steps, with another non-electrochemical Tafel step (H* + H* → $H_2$) being investigated in this work as well (**SI Section 11**). Researchers have done extensive theoretical studies on PCET steps in HER, typically by employing the CHE model and the charge extrapolation scheme.[27–29] The exact mechanism and quantitative kinetic properties of the above PCET steps are still under debate, with discrepancies of activation energies between the computational predictions and experimental results.[30–33] A recent theoretical work using the CMD method tried to reconcile the above-mentioned discrepancies,[12] but there are still unresolved issues.

With the NQEs being explicitly considered in this study, we attempt to provide new insights into the impact of this quantum effect on the PCET steps involved in electrocatalytic HER. Our simulations reveal that proton tunneling exhibits a substantial impact on the free energy profile of elementary PCET reactions in HER, and more importantly, provides new physical pictures for understanding pathways of the transferring proton overcoming kinetic barriers in the Volmer and Heyrovsky steps. We will first introduce the main idea and principles of our developed computational algorithm, and then present the major results elucidating the NQEs in electrocatalytic HER in the following sections.

## Results
### Methodology principles

Firstly, we introduce the principle of the GC-HMC algorithm proposed by this study. Considering a system with the coordinates of all particles and the total electron number represented by $\vec{R}$ and $N_e$, respectively, the GC ensemble partition function can be expressed as

$$\Xi(\beta, \mu_e) = \sum_{N_e} \Lambda \int d\vec{R} \exp[-\beta(E(\vec{R}, N_e) - \mu_e N_e)] = \sum_{N_e} \exp(\beta \mu_e N_e) Q(\beta, N_e) \quad (1)$$

where $\beta = \frac{1}{k_B T}$ is the inverse temperature, $\Lambda$ is the prefactor generated by the integral of momenta degrees of freedom, $E(\vec{R}, N_e)$ and $\mu_e$ express the potential energy and the electrochemical potential of the external electronic reservoir in equilibrium with the system respectively, and $Q(\beta, N_e)$ refers to the canonical ensemble partition function (integral in the phase space) at a specific temperature $\beta$ and with a total number of electrons $N_e$. For the exact GC ensemble condition, the work function of an instantaneous configuration will fluctuate around -$\mu_e$ during the simulation, rather than being fixed at -$\mu_e$ which corresponds to the situation using the GC-DFT method.

Sufficient configurational sampling of complex electrode/solvent interfacial environment is essential for studying PCET steps. Due to the issue of particle-number variations in GC conditions, MC[34,35] is a more practical method for open system simulations with particle insertion/deletion compared to MD.[35,36] We employ the HMC[37,38] method to improve the sampling efficiency with multi-particle displacements. Free energy profiles provide important physical insights into the thermodynamic and kinetic properties of PCET steps. The thermodynamic integration (TI) method[7,8,39] is appropriate for computing the free energy profiles with constraints on a defined RC. Considering the above requirements, a constrained HMC approach proposed in our previous study[40] is employed in this work. The details of the constrained HMC method are shown in the **Methods** section.

To achieve an exact constant potential condition, the constrained HMC method can be easily extended to a constrained GC-HMC method by incorporating an extra degree of freedom – the total electron number $N_e$ of the surface models. We can sample $N_e$ by the Metropolis algorithm with particle coordinates fixed, and the corresponding acceptance probability is

$$A(N_e'|N_e) = \min\left[1, \exp\left(-\beta\left(E(\vec{R}, N_e') - E(\vec{R}, N_e) - \mu_e(N_e' - N_e)\right)\right)\right]$$

(2)

For the exploration of particles' positions, we employ the constrained HMC method[40] at a fixed $N_e$ value. The workflow of the GC-HMC algorithm is shown in **Figure 1(a)**. A MC sampling trajectory begins with an initial structure, characterized by the composite configuration $[\vec{R}, N_e]$ of the system. Three types of degrees of freedom are considered: the total number of electrons ($N_e$), the centroid of atomic coordinates ($\vec{R}$), and internal degrees of freedom within the quantized beads' configurations ($\vec{R}_\Delta^{(k)}$). We then randomly select which type of degrees of freedom to be perturbed by trial moves at each MC step, based on a random number $\xi$ satisfying a uniform distribution within $[0, 1]$. We vary the total electron number $N_e$ based on **Eq. 2** to achieve a GC constant potential condition, or perform HMC for the centroid of atomic positions ($\vec{R}$) to sufficiently sample the complex interfacial structure, or use the PIMC method to treat NQEs, by which we update $\vec{R}_\Delta^{(k)}$ based on the staging algorithm.[55] Estimators of targeted physical quantities are subsequently evaluated. We repeat the above process until reaching the required total MC step number, and finally obtain the ensemble average of interested physical quantities.

In fact, a potentiostat-based constant potential MD algorithm was proposed in 2012[41], which is also a sampling approach with instantaneous work function fluctuations conforming with an exact grand canonical distribution, equivalent to the GC-HMC method proposed in our study. If we aim to accomplish a constant pH condition in future, our GC-HMC algorithm is easier to accommodate to variable proton-number sampling than the forementioned MD algorithm[41] which has to deal with the discontinuity issue of particle insertion/deletion. To fulfill a constant pH condition, the number of protons in the electrode/solvent system needs to be dynamically adjusted during sampling. This is similar to the strategy for maintaining a constant potential by varying $N_e$ as discussed above, which could be implemented as an independent module in the computational framework (**Figure 1(a)**) in our future study.

Since the NQEs play an important role even at room temperature, the Feynman PI algorithm[23,24] is employed to consider the quantum feature of protons in the PCET steps. The quantum GC ensemble partition function can be expressed as

$$\Xi_{\text{qtm}}(\beta, \mu_e) = \sum_{N_e} \exp(\beta \mu_e N_e) \, Q_{\text{qtm}}(\beta, N_e)$$

(3)

where $Q_{\text{qtm}}(\beta, N_e)$ refers to the quantum canonical ensemble partition function. The expression of the quantum GC ensemble partition function $\Xi_{\text{qtm}}(\beta, \mu_e)$ is analogous to the classical one shown in **Eq. 1**, please refer to **SI Section 1** for detailed derivation of $\Xi_{\text{qtm}}(\beta, \mu_e)$. Combining the PIMC algorithm with the GC-HMC method, the quantum effect can be taken into account together with sufficient configurational sampling of the electrode/solvent interface. The details of the PIMC algorithm implemented in this study are shown in the **Methods** section. In the quantum case, the expression of the acceptance ratio of the total electron number $N_e$ trial move is similar to **Eq. 2**, except that the potential energy $E(\vec{R}, N_e)$ in **Eq. 2** is replaced by the average potential energy of all beads sharing the same electron number $N_e$ in the PI formalism (refer to **SI Section 1** for detailed discussions).

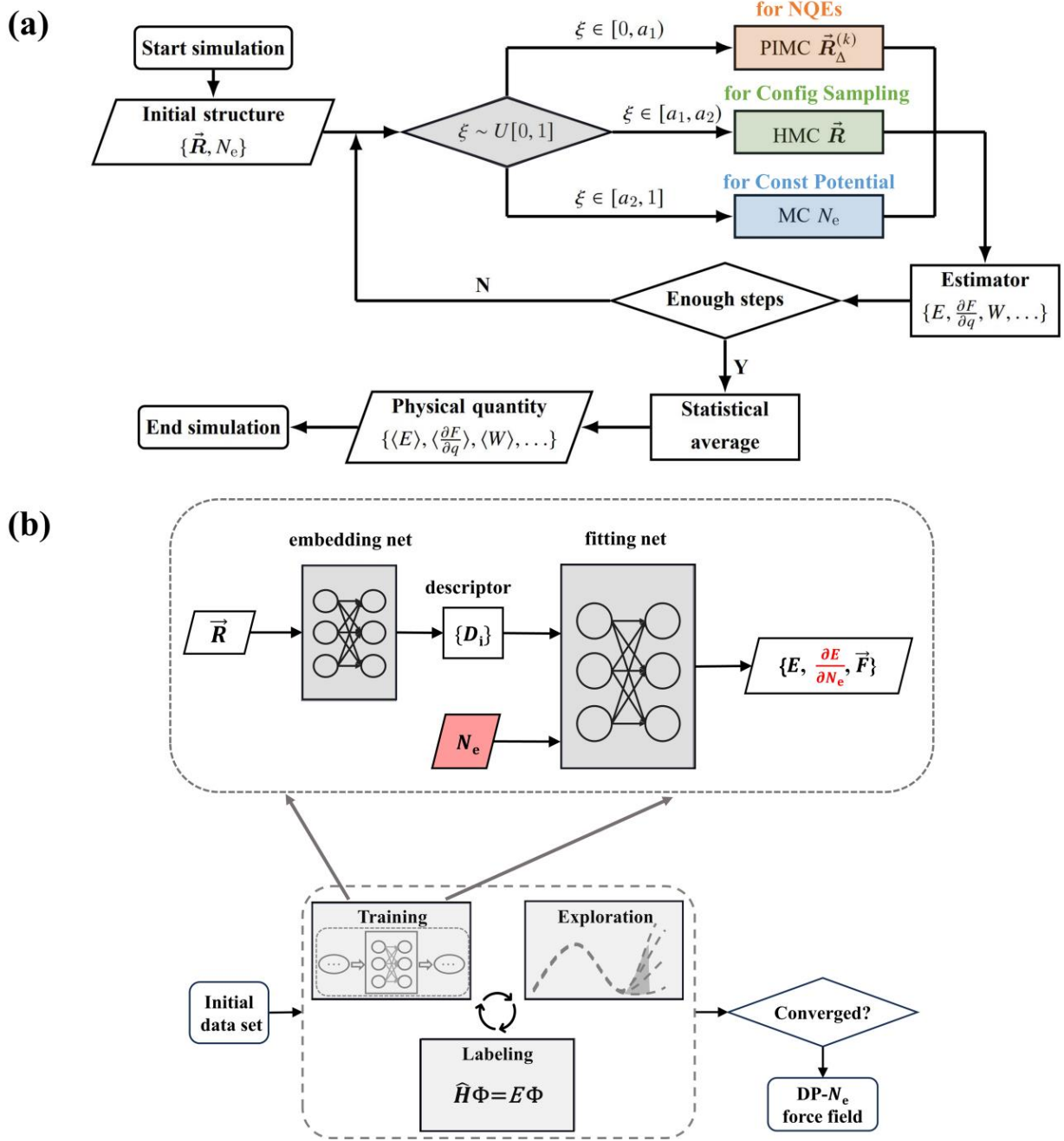

**Figure 1.** (a) Workflow of the GC-(PI)HMC method. Trial moves for different types of degrees of freedom are selected based on a preset ratio using a random variable $\xi$ satisfying a uniform distribution within $[0, 1]$, and the probabilities of making trial moves for the internal degrees of freedom within the quantized beads' configurations in PIMC ($\vec{R}_\Delta^{(k)}$), centroid atomic coordinates ($\vec{R}$), total number of electrons ($N_e$) are $a_1, a_2 - a_1, 1 - a_2$, respectively. (b) Construction framework and training workflow of the DP-$N_e$ MLP adopted in this work. Initially, a data set is provided, followed by an iterative process which automatically goes through training, exploration and labeling steps. The iteration is considered converged after the accurate-sample percentage among the newly explored configurations is above 85%. The zoom-in schematic plot above the training workflow illustrates the construction framework of the DP-$N_e$ force field. Atomic coordinates $\vec{R}$ of a modelling system are inputs for the embedding network generating descriptors $\{D_i\}$. The fitting network maps $\{D_i\}$ together with an extra degree of freedom $N_e$ to the total energy $E$, atomic forces $\{\vec{F}_i\}$, and $\partial E / \partial N_e$ which relates to the work function of this extended configuration [$\vec{R}$, $N_e$] as discussed in following sections.

**Framework of our developed DP-$N_e$**

A traditional MLP model, such as DP,[25] typically substitutes atomic coordinates as inputs and infer the total energy and atomic forces of a modeling system. In this work, we introduce a new degree of freedom to input parameters: the total number of electrons $N_e$ of our modeling interfacial system (**Figure 1(b)**). To enable computing the work function ($W$) of an instantaneous sampled configuration in the extended space [$\vec{R}$,$N_e$], we include an additional output $\frac{\partial E(\vec{R},N_e)}{\partial N_e} = -W(\vec{R},N_e)$. For a valid GC ensemble sampling, it is essential to satisfy:

$$< \frac{\partial E(\vec{R},N_e)}{\partial N_e} >_{GC} = < -W(\vec{R},N_e) >_{GC} = \mu_e \qquad (4)$$

where $<\cdot>_{GC}$ denotes the statistical GC ensemble average, details of which can be referred to the following results and **SI Section 4**. **Figure 1(b)** shows the general framework of the DP-$N_e$ model adopted in this study. Our developed DP-$N_e$ MLP facilitates the sampling of GC ensembles with variable electron numbers. Chen et al. proposed a practical approach for machine-learning emulation derived from the GC-DFT method,[14] which incorporates the electrode potential as a new degree of freedom into the input parameters. However, the construction of this machine-learning force field actually encounters the same issue as the GC-DFT approach, which does not perform an exact sampling of the microstates' distribution based on the GC partition function, but instead enforces a fixed work function constraint.

We validate the accuracy of our DP-$N_e$ MLP by comparing the inferred energies and forces with the DFT results, and a good agreement on the testing dataset is achieved as shown in **Figure S3**. The total root mean square errors (RMSEs) of energies and forces on the testing dataset are 0.6 meV/atom and 57 meV/Å for configurations along the Volmer reaction path, and 1.1 meV/atom and 71 meV/Å for the Heyrovsky reaction case. Such small errors indicate the reliability of our DP-$N_e$ force field model in describing PES of HER steps with respect to the extended degrees of freedom [$\vec{R}$, $N_e$].

**Construction of the atomic interface model**

We study the PCET steps involved in HER on a (5×5) Pt (111) surface slab composed of four atomic layers illustrated in **Figure 2(a)**. The modeled electrode surface contains 100 Pt atoms with 1 monolayer (ML) hydrogen coverage at Pt atop sites.[42,43] We also include two water layers (36 explicit water molecules) to take account of the solvation effect and provide proton donors. A vacuum region of 15 Å thickness is further added above the water layers in order to decouple periodic images of the slab model. Since we adjust the total electron number of the interface system in our GC-HMC algorithm, compensating charge has to be included to maintain the overall charge neutrality of the supercell under a periodic boundary condition (PBC). We employ the scheme of placing a compensating charge plate in the vacuum region right above the water solvation layer to mimic an effective electric double layer at the electrode/solution interface, analogous strategies can be found in earlier theoretical work.[28,44,45] We realized the above function in the first-principles package Atomic-orbital Based Ab-initio Computation at USTC (ABACUS)[46,47] used in this study, consistent with the corresponding algorithm implementation in the popular DFT code Quantum ESPRESSO (QE).[48,49] We further benchmark our calculated electrostatic energy profiles with the QE results on a testing interface model. Details can be found in **SI Section 2**.

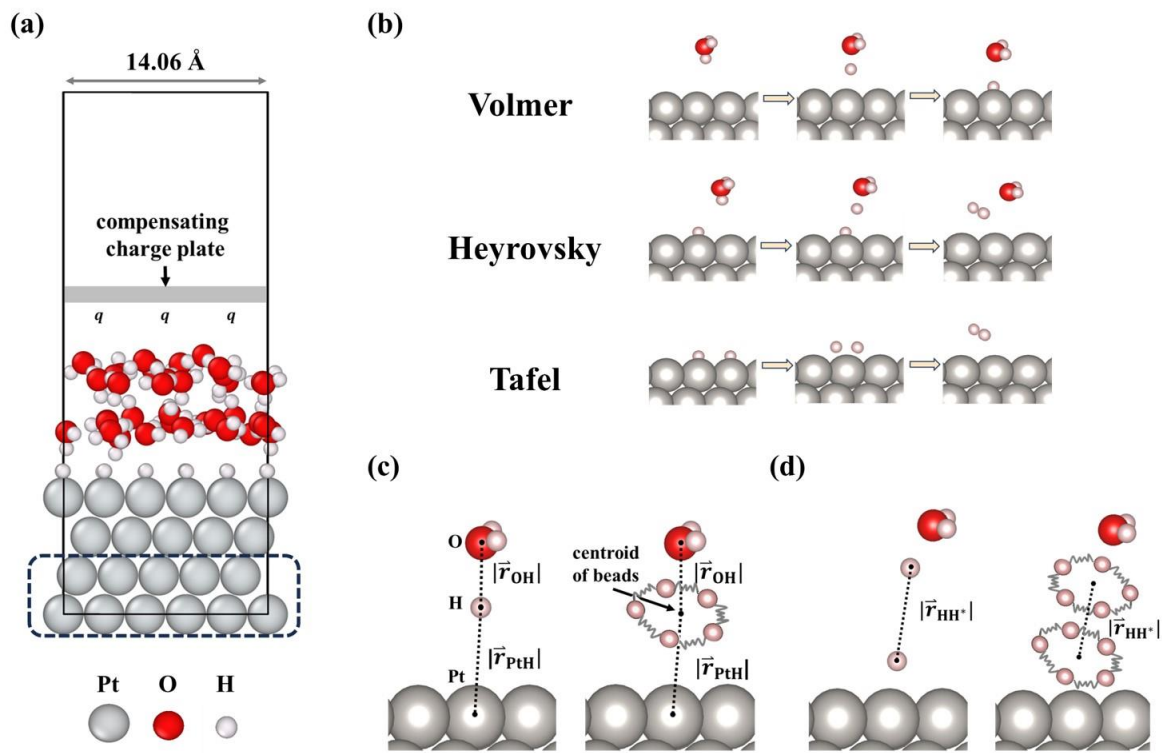

**Figure 2.** (a) Side view of the simulated (5×5) Pt (111) surface slab composed of four atomic layers. The interface model contains a water bilayer over a 1 ML of adsorbed hydrogen. A compensating charge plate is placed in the vacuum region above the water solvation layer. The Pt atoms inside the black dashed rectangle are fixed along all simulation trajectories. (b) Schematic diagrams of the proton transfer pathways of three elementary reactions (Volmer, Heyrovsky, Tafel) involved in HER. Illustration of the key atoms relevant to the RC definitions of the (c) Volmer and (d) Heyrovsky steps for classical (left) and quantum (right) cases. Adsorbed hydrogen atoms and water molecules not directly participating in the investigated reactions are not displayed in these schematic plots for clarity.

The HER process starts with a Volmer reaction, which is a PCET step of a proton's adsorption forming an H*. Following this, the evolution of an $H_2$ molecule can proceed along two distinct pathways: the Volmer-Tafel or the Volmer-Heyrovsky mechanisms. The proton transfer pathways of the Volmer, Heyrovsky and Tafel steps are illustrated in **Figure 2(b)**. We note that the Tafel step, characterized as a chemical (instead of an electrochemical) reaction with negligible charge transfer across the interface, is beyond the scope of PCET steps, thus is not extensively discussed in the main text. We present our investigation of the NQEs on the Tafel reaction in **SI Section 11**.

We need to define reasonable RCs for different types of reaction paths, so that we could drive reactions and perform the mean force integration by our GC-(PI)HMC. In the present work, we employ the difference of bond distances $|\vec{r}_{\text{PtH}}|$ and $|\vec{r}_{\text{OH}}|$ as the RC $q_{\text{Volmer}}$ for the Volmer reaction (**Eq. 5**), in which $|\vec{r}_{\text{PtH}}|$ is the distance between the Pt atom for H adsorption and the transferring proton, and $|\vec{r}_{\text{OH}}|$ is the distance between the proton donating oxygen (belonging to an H$_2$O molecule above the Pt site) and the transferring proton (illustrated in the left panel of **Figure 2(c)**). For the Heyrovsky reaction, we define the RC $q_{\text{Heyrovsky}}$ as the distance $|\vec{r}_{\text{HH}^*}|$ between the two hydrogen atoms forming the H$_2$ molecule (**Eq. 6**), the definition is schematically depicted in the left panel of **Figure 2(d)**.

Volmer RC: $$q_{\text{Volmer}} = |\vec{r}_{\text{PtH}}| - |\vec{r}_{\text{OH}}| \tag{5}$$

Heyrovsky RC: $$q_{\text{Heyrovsky}} = |\vec{r}_{\text{HH}^*}| \tag{6}$$

In the quantum situations for studying NQEs, the transferring H in the Volmer step and the two combining H atoms in the Heyrovsky step are quantized as beads configurations isomorphic to a ring-polymer model,[16,17,19,50] we then treat the centroid of the corresponding ring-polymer beads (illustrated in the right panels of **Figure 2(c) and 2(d)**) as the positions of the quantized H atoms.[17,19]

**Statistical results of potential and charge variations from our GC sampling**

Since the PCET steps involve electron transfer across interfaces, the activation energies of Volmer and Heyrovsky steps have a dependence on applied potentials. It is through the work function that we can establish a connection between the reduction voltage $U$ relative to the standard hydrogen electrode ($\varphi_{\text{SHE}}$, ~ 4.4 V vs the vacuum level[51,52]) as follows:

$$U = \frac{W}{|e|} - \varphi_{\text{SHE}} \tag{7}$$

where $|e|$ means the unit charge, and the work function $W(\vec{R},N_e)$ of an instantaneous microstate, sampled in our GC-(PI)HMC trajectory within the extended $[\vec{R},N_e]$ configurational space, can be obtained from $\frac{\partial E(\vec{R},N_e)}{\partial N_e}$ as introduced in the above section (refer to **SI Section 4** for detailed DFT validations). We are able to achieve different electrochemical reducing conditions in the GC sampling for this open system by adjusting the electrochemical potential parameter $\mu_e$ of the external electronic reservoir (shown in **Eq. 1**), equivalent to the potentiostat scheme proposed by an earlier theoretical work.[41] We thus can compute free energy profiles of the investigated PCET steps at different applied potentials. **Figure 3(a)** shows that the GC ensemble average <

$W(\vec{R},N_e) >_{GC}$ equals to the controlling parameter $-\mu_e$, which is consistent with **Eq. 4** and justifies the validity of our GC sampling. The magnitude of the work function fluctuation range is also consistent with the earlier results obtained by the above-mentioned potentiostat scheme (~ ±0.5 eV).[41] The modeled system corresponds to the Volmer step with $q_{Volmer} = 0.21$ Å at $\mu_e$ = -3.5 eV vs vacuum ($U$ = -0.9 V vs SHE).

The modeled system's total electron number $N_e$ also fluctuates around an average value as shown in **Figure 3(b)** and exhibits a normal distribution **(SI Section 5)** satisfying the GC ensemble distribution. We use the extra electron number ($N_e^{extra}$) added/subtracted to/from the Pt/H$_2$O interface model to represent the total electron number for clarity. The average value $< N_e >_{GC}$ corresponds to the system's charge state at a specific applied potential condition. We need to emphasize that an exact thermodynamic simulation of an electrochemical open system should obey the fundamental principle of the GC ensemble distribution (**Eq. 1**), where $N_e$ and $W(\vec{R},N_e)$ are a pair of conjugate thermodynamic variables of a microstate. A correction formulation of GC sampling should exhibit the feature that neither $N_e$ nor $W$ of sampled microstates is fixed along a simulation trajectory, while the ensemble average $< W(\vec{R},N_e) >_{GC}$ equals to the controlling parameter $-\mu_e$ of an external electronic reservoir. An analogous concept applies to the case of an isothermal-isobaric *NPT* ensemble sampling, where neither volume $V$ nor pressure $P(\vec{R},V)$ of the sampled microstates is fixed, while the average pressure $< P(\vec{R},V) >_{NPT}$ equals to the setup external pressure. The above fundamental principle is rarely treated exactly in the electrocatalytic simulation community, where fixed charge or fixed potential schemes were commonly adopted in the potential energy or free energy calculations for electrochemical PCET steps.

A gradual increase of total electron number is observed from the IS to the FS along the reaction pathways of both the Volmer and Heyrovsky mechanisms (shown in **Figure 3(c)**), which is as expected because the electrode slab needs to keep acquiring electrons to facilitate the progress of reduction reactions when in equilibrium with an electronic reservoir at a constant electrochemical potential. We further conduct detailed calculations to examine the relationship of charge states of the Volmer step's IS and FS with applied potentials **(SI, Section 6)**. The results exhibit a consistent trend that a more negative potential leads to larger electron numbers in the modeled system, indicating a reasonable electronic response in PCET reactions that a stronger reducing driving force requires a higher concentration of electrons in the system.

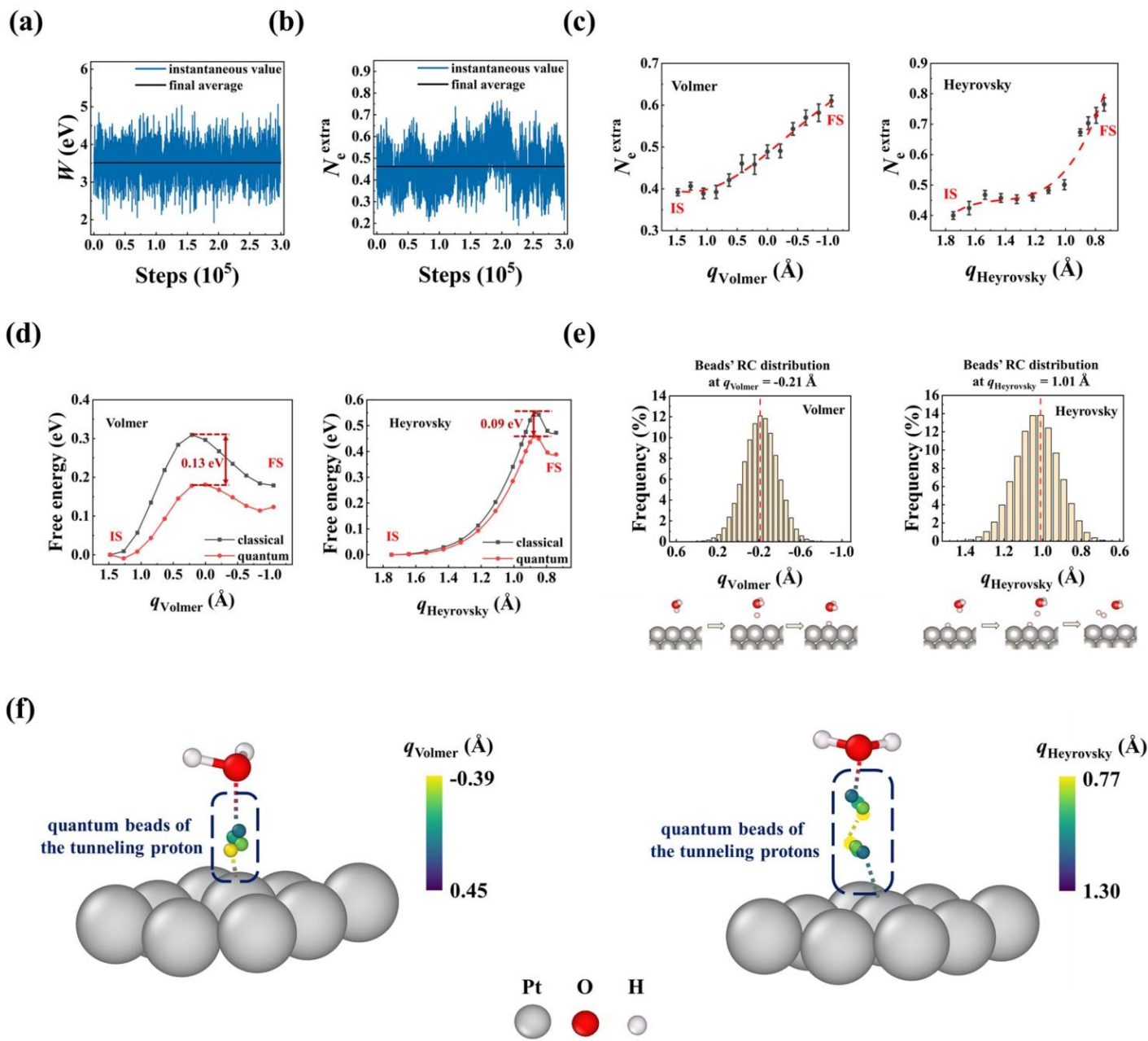

**Figure 3**. (a) Work function and (b) total electron number fluctuation with respect to MC steps of the Volmer reaction with a RC fixed at $q_{\text{Volmer}} = 0.21$ Å. The extra electron number added/subtracted to/from the model ($N_e^{\text{extra}}$) is used to represent the total electron number for clarity. Blue curves show the instantaneous values, and black lines are the final averages along complete sampling trajectories. (c) Total electron number (represented by $N_e^{\text{extra}}$) change along the reaction path from IS to FS of classical Volmer and Heyrovsky simulations. Error bars represent standard errors from 15 (Volmer) or 10 (Heyrovsky) independent simulations. Red dashed curves are simply guidelines showing the trend of electron number

variation with respect to RC. (d) Classical and quantum free energy profiles with respect to the defined RC of the Volmer and Heyrovsky PCET steps. (e) RC distribution associated with the ring-polymer beads in the PI simulations of Volmer and Heyrovsky reactions. Red dashed vertical lines are the constrained RC values defined for the proton beads' centroid in our PIMC, which also correspond to the constrained RC values in classical simulations without RC spreading. (f) Structural plots explicitly showing the spreading beads of the transferring protons or hydrogen atoms at $q_{\text{Volmer}} = 0.0$ Å for the Volmer reaction and $q_{\text{Heyrovsky}} = 0.95$ Å for the Heyrovsky reaction. Adsorbed hydrogen atoms and Pt layers not directly participating in the studied reactions are not displayed here, and only one water molecule (the proton donor) remains in the plot for clarity. Only part of the 16 ring-polymer beads are shown. All results in this figure are obtained under a specific reduction potential of $U = -0.9$ V vs SHE (or $\mu_e = -3.5$ eV vs vacuum) at $T = 300$ K.

**Impact of NQEs on activation energies of PCET steps in HER**

We obtain the free energy profiles of the Volmer and Heyrovsky PCET steps at $\mu_e = -3.5$ eV vs vacuum ($U = -0.9$ V vs SHE), presented in **Figure 3(d),** by numerically integrating the mean forces at different RCs along the reaction paths.[19,40,53] To further investigate the impact of NQEs on these PCET free energies, we implement the PIMC algorithm into the GC-HMC method (illustrated by the workflow plot in **Figure 1(a)**). We then compare the free energy results between the classical and quantum situations. The computational details of our sampling calculations are presented in **SI Section 8**. The quantitative influence on the activation energies ($E_a$) of the Volmer and Heyrovsky steps are 0.13 eV and 0.09 eV respectively (at $U = -0.9$ V vs SHE) upon incorporating NQEs of the transferring protons in these PCET catalytic steps (**Figure 3(d)**). When considering NQEs in the free energy calculations, the predicted activation energies exhibit a notable decrease, leading to a non-negligible enhancement of the investigated PCET reaction rates compared to the results of classical cases. The above results indicate that reaction rates could be underestimated by approximately 50-100 folds at $T = 300$ K if we employ the traditional view of treating protons as classical particles, based on the Arrhenius type equation $r \propto e^{-\frac{E_a}{k_B T}}$.

We realize some readers may raise a question here that, since DFT calculations with different setup typically exhibit an uncertainty or deviation range of about 0.1 – 0.2 eV in describing energetics of electrocatalytic steps, is the reported quantitative NQEs' impact of 0.1 – 0.15 eV (shown in **Figure 3(d) and Figure 4**) on PCET activation energies reliable or of physical meaning in this work? The basic logic of our response is that, as we keep all of the first-principles calculations consistent within this study and our adopted DFT setup also follows a routine of simulations in electrocatalysis, the uncertainty caused by inconsistent first-principles calculation setup can be excluded, meaning that the important qualitative impact of NQEs on HER PCET steps is expected to persist if other *ab-initio* setup schemes are employed. We actually find that earlier theoretical predictions (all performed in a classical situation) of the PCET activation energies in electrocatalytic HER at Pt surfaces were consistently higher than those from experimental measurements.[30–33] The NQEs thus could be a key factor for reconciling the above discrepancy, as revealed by our computational study which explicitly deals with the NQEs under an exact GC ensemble sampling.

Proton tunneling originates from the intrinsic quantum nature, which is already demonstrated to be remarkable even at room temperature for the HER PCET steps. To achieve a clearer and qualitative understanding of the transferring proton's quantum feature, we analyze the quantum beads expansion for the states at near-TS RC in the Volmer path ($q_{\text{Volmer}}$ = -0.21 Å) and the Heyrovsky path ($q_{\text{Heyrovsky}}$ = 1.01 Å) at $U$ = -0.9 V vs SHE in **Figure 3(e)**. Let's first consider in a classical picture, the positions of the transferring protons or hydrogen atoms are essentially mass points obeying the constraint of a specific RC. If we plot the RC value of each sampled configuration in a GC-HMC trajectory, all of which must fall onto a single value, corresponding to the red dashed vertical lines in **Figure 3(e)**. However, the situation is quite different in the quantum case, since a proton or hydrogen atom can "split" into multiple beads in the PI algorithm, the "uncertainty" of a micro particle's position just reflects its quantum feature. Therefore, if we do the statistics of each bead configuration's RC value along our GC-PIHMC sampling trajectory and plot the histogram (**Figure 3(e)**), the RC distribution would spread around an average number (actually equals to the constrained RC value defined for the proton beads' centroid). We can clearly see that the quantum-treated transferring proton exhibits an IS-TS-FS mixed feature at the TS of the Volmer or the Heyrovsky path, as illustrated by the schematic atomic structures underneath the histogram plots (**Figure 3(e)**). This observation indicates a sharp contrast with the classical deterministic understanding of transferring protons along PCET paths, and further explains the lowered activation energy due to the tunneling effect. This is because the TS in the PIMC simulations also mixes with ring-polymer beads with RCs corresponding to IS and FS, which have lower potential energies than the classical TS configurations.

A more direct configurational visualization of the spreading beads of the transferring protons or hydrogen atoms in the Volmer and Heyrovsky path are presented in **Figure 3(f)**. We conduct a quantum PI sampling at a specific near-TS configuration where the coordinates of all classical particles and the centroids of the quantized H are fixed, with the relative positions of the quantized H ring-polymer beads (or considered as the internal degrees of freedom within a quantized particle) being sufficiently sampled by the PI algorithm. We randomly choose a specific configuration of the spreading beads and show the structure plots in **Figure 3(f)**, which exhibits a considerable uncertainty in the quantized proton's or H atom's position and an unusual feature that IS and FS configurations are mixed in the TS sampling by the PI simulations. For example, the IS-like configuration ($H^+_{sol} + H^*$) denoted by the blue color and the FS-like configuration ($H_2$) highlighted by the yellow color coexist in the Heyrovsky TS PIMC sampling (the right panel of **Figure 3(f)**), resulting in an emergence of the tunneling behavior (note that we only show part of the 16 ring-polymer beads for clarity). We thus can obtain a qualitative physical picture that the transferring proton or H atom exhibits wave-like quantum characteristic, facilitating the particles to tunnel through classical barriers along the PCET pathways in HER, leading to a remarkable activation energy reduction compared to the classical simulations. The NQEs revealed by our theoretical work thus contribute new physical insights into the fundamental understanding of PCET dynamics in electrocatalytic HER.

**Insights into the HER mechanism inspired by the consideration of NQEs**

The above results reveal the exotic quantum behavior of the transferring protons during PCET processes. We are now interested in the impact of NQEs on our understanding of the electrocatalytic HER mechanism. Two well-known reaction pathways compete with each other, that is, the Volmer-Heyrovsky pathway against the Volmer-Tafel pathway. **Figure 4** shows the activation energies of the Volmer, Heyrovsky, and Tafel elementary steps with respect to applied electrochemical potentials for both of the classical and quantum situations. We note that the activation energy of the non-electrochemical Tafel step is expected to exhibit negligible dependence on applied potentials, we thus plot its activation energy as a constant value (derived from our calculations with details shown in **SI Section 11**) in **Figure 4**. Our classical activation energy results are in good agreement with a recent computational work[12], where similar activation energies of both the Tafel step (0.53 eV) and the Volmer step (0.25 - 0.50 eV) were reported in their study within the relevant voltage range investigated in this work, justifying our GC-HMC calculations under the classical situation. If the transferring proton is treated as a classical particle, the Volmer-Tafel path is more likely to dominate the $H_2$ production under the electrochemical condition of voltage $U \geq$ -0.9 V vs SHE. However, due to the proton tunneling behavior resulting from its intrinsic quantum nature, more exact activation energies of the Heyrovsky step considering NQEs are ~ 0.1 eV lower than those derived from the classical cases. Since the Tafel steps are almost not affected by NQEs (discussed in **SI Section 11**), we can see from **Figure 4** that the transition point from the Volmer-Tafel path to the Volmer-Heyrovsky path is significantly shifted to a less reducing potential region by a difference of 0.5 V, indicating a tendency of the electrochemical Heyrovsky step suppressing the chemical Tafel step toward a smaller overpotential condition. This observation from our work matches with an experimental-theoretical-joint analysis[54] claiming dominancy of the Heyrovsky step across most of the relevant electro-reducing potential range. Furthermore, our results provide possible explanations to a perplexity raised by Kronberg et al.,[12] in which only the Volmer and Tafel steps were simulated by constrained MD simulations, that the Volmer-Tafel mechanism cannot recover the Tafel slope observed in experiments.[54] It is highly likely that the Volmer-Heyrovsky path suppresses the alternative path when considering the NQEs. At the end, we also need to point out a possible limitation inherent in the mean-force integration method for free energy profile calculations under constrained sampling algorithms. Since the solvent structures are sampled independently at each RC along the reaction pathway, our constrained HMC approach may overestimate the solvent reorganizations leading to a softened potential dependence of PCET activation energies in **Figure 4**.[12]

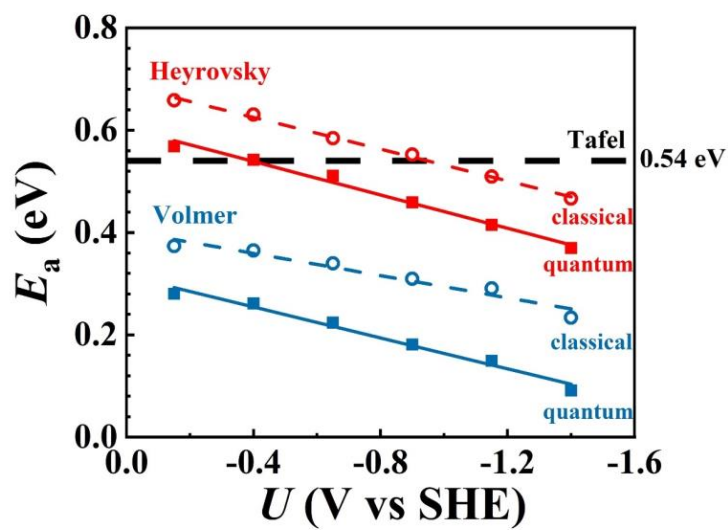

**Figure 4**. Activation energy results of Volmer and Heyrovsky mechanisms with respect to electrochemical potentials under classical and quantum situations at $T$ = 300 K. Blue (Red) hollow circles and solid squares represent the Volmer (Heyrovsky) reaction for the classical and quantum cases, respectively. Solid and dashed lines derived from a linear fitting simply perform as guidelines illustrating the trend. The black dashed line represents the activation energy result (0.54 eV) of the Tafel reaction, which does not change with respect to voltage variation, and has almost identical values under classical and quantum conditions.

## Discussion

In this work, we employ our proposed GC-(PI)HMC sampling framework, assisted by a MLP adapted for electrochemistry, to reveal a non-negligible role of NQEs in computational work to achieve an exact quantitative description of electrochemical mechanisms, which is also consistent with earlier studies reviewed at the beginning of this article that NQEs exhibit remarkable influence in many chemical/materials systems[15–19] even at room temperature.

On the basis of this MC framework, we emphasize that the good expandability of our sampling workflow provides convenience for incorporating more physical factors in future code development. For example, the MC sampling for proton number variation, to achieve a constant pH condition, can be implemented as a new sampling branch in the workflow shown in **Figure 1(a)**, which is an independent module and does not affect the programs of the other sampling functions for different types of degrees of freedom.

More importantly, the physical insight of proton tunneling provided by our simulations based on an exact constant potential GC sampling can be extended to broader scenarios in electrocatalysis, including ORR, $CO_2$RR or nitrogen reduction reaction (NRR) systems. We may find similar qualitative NQEs impact on the kinetics of elementary PCET steps, which may reshape our fundamental understanding on electrocatalytic reactions. Our computational work thus highlights the importance of incorporating protons' NQEs in modeling PCET steps existing ubiquitously in energy conversion systems.

At the end, we need to point out a few limitations of our current computational framework. First, the thermodynamic integration method used in our approach produces free energy profiles, which are just thermodynamic results and lack realistic dynamic information. Second, the adiabatic approximation employed in this work ignores the non-adiabatic effect for now, which could be an important factor worthy of investigation in certain PCET reactions.

## Methods

### Methodology of the constrained HMC algorithm and derivations

For a system with all particle positions $\vec{R}$ and potential energy $U(\vec{R})$, we define a RC $q = f(\vec{R})$ to monitor the reaction process, and the reaction mechanism can be explored via MC simulations at different fixed $q$ values using the constrained HMC algorithm.[40] The key idea of the algorithm is to integrate the RC after a coordinate transformation and sample the rest two types of degrees of freedom with the corresponding two types of MC schemes. The procedures of the algorithm can be divided into the following three steps:

a. Apply an appropriate coordinate transformation on $\vec{R}$ to obtain general coordinates $\vec{q} = (q, \vec{q}_{\text{trans}}, \vec{q}_{\text{primit}})$, where $\vec{q}_{\text{trans}}$ and $\vec{q}_{\text{primit}}$ represent the transformed coordinates related to the RC and the primitive coordinates maintained the same in the transformation, respectively. Then we integrate the RC in the probability density of $q = s$ as follows

$$P(s) = \frac{\Lambda}{Q(\beta, N_e)} \int d\vec{R} \exp[-\beta E(\vec{R}, N_e)] \, \delta(q - s)$$

$$= \frac{\Lambda}{Q(\beta, N_e)} \int d\vec{q} \exp[-\beta \tilde{E}(\vec{q}, N_e)] \, \delta(q - s) \mathcal{J}(\vec{q})$$

$$= \frac{\Lambda}{Q(\beta, N_e)} \int \mathrm{d}q \mathrm{d}\vec{q}_{\text{trans}} \mathrm{d}\vec{q}_{\text{primit}} \exp[-\beta \tilde{E}(\vec{q}, N_e)] \delta(q-s) \mathcal{J}(q, \vec{q}_{\text{trans}})$$

$$= \frac{\Lambda}{Q(\beta, N_e)} \int \mathrm{d}\vec{q}_{\text{trans}} \mathcal{J}(s, \vec{q}_{\text{trans}}) \int \mathrm{d}\vec{q}_{\text{primit}} \exp[-\beta \tilde{E}(s, \vec{q}_{\text{trans}}, \vec{q}_{\text{primit}}, N_e)]$$

(8)

where $\mathcal{J}(\vec{q})$ is the Jacobian related to the coordinate transformation and $\mathcal{J}(\vec{q}) = \mathcal{J}(q, \vec{q}_{\text{trans}})$.

b. The primitive coordinates $\vec{q}_{\text{primit}}$ together with the particle momenta $\vec{p}_{\text{primit}}$ are combined to obtain the partial Hamiltonian $\mathcal{H}(\vec{p}_{\text{primit}}, \vec{q}_{\text{primit}})$ used in the HMC scheme.[37,38]

c. Select the sampling type of degrees of freedom at random based on the preset ratio and sample these degrees of freedom with the rest ones fixed. Specifically, $\vec{q}_{\text{trans}}$ evolves via the usual Metropolis scheme and the acceptance probability is

$$A(\vec{q}'_{\text{trans}}|\vec{q}_{\text{trans}}) = \min\left\{1, \frac{\mathcal{J}(s, \vec{q}'_{\text{trans}})}{\mathcal{J}(s, \vec{q}_{\text{trans}})} \exp\left[-\beta\left(\tilde{E}(s, \vec{q}'_{\text{trans}}, \vec{q}_{\text{primit}}, N_e) - \tilde{E}(s, \vec{q}_{\text{trans}}, \vec{q}_{\text{primit}}, N_e)\right)\right]\right\}$$

(9)

and $\vec{q}_{\text{primit}}$ evolves via the HMC scheme and the acceptance probability is

$$A(\vec{q}'_{\text{primit}}|\vec{q}_{\text{primit}}) = \min\left\{1, \exp\left[-\beta\left(\mathcal{H}(\vec{p}'_{\text{primit}}, \vec{q}'_{\text{primit}}) - \mathcal{H}(\vec{p}_{\text{primit}}, \vec{q}_{\text{primit}})\right)\right]\right\}$$

(10)

The workflow of the constrained HMC method is shown in **Figure 1(a)** and more details can be referred to our earlier theoretical work.[40]

**Methodology of the PI algorithm and derivations**

Consider the same system in the above section, we represent the masses of all particles with a diagonal matrix $\underline{m}$. In quantum mechanics, the partition function of a system is expressed as the trace of a density operator, which can be derived as follows using a factorization formalism proposed by Feynman[23,24]

$$Q_{\text{qtm}}(\beta, N_e) = \text{Tr}[\exp(-\beta \hat{\mathcal{H}})]$$

$$= \lim_{P \to \infty} \text{Tr}\left[\exp(-\beta \hat{\mathcal{H}}/P)^P\right]$$

$$= \lim_{P \to \infty} \int \mathrm{d}\vec{R}^{(1)} \ldots \mathrm{d}\vec{R}^{(P)} \langle \vec{R}^{(1)}|\exp(-\beta \hat{\mathcal{H}}/P)|\vec{R}^{(2)}\rangle$$

$$\times \cdots \times \langle \vec{R}^{(P-1)}|\exp(-\beta \hat{\mathcal{H}}/P)|\vec{R}^{(P)}\rangle \langle \vec{R}^{(P)}|\exp(-\beta \hat{\mathcal{H}}/P)|\vec{R}^{(1)}\rangle$$

$$= \lim_{P\to\infty} \Lambda_P \int d\vec{R}^{(1)} \ldots d\vec{R}^{(P)} \exp\left\{-\beta \sum_{k=1}^{P} \left[\frac{1}{2}\omega_P^2 \left(\vec{R}^{(k+1)} - \vec{R}^{(k)}\right)^T \underline{m}\left(\vec{R}^{(k+1)} - \vec{R}^{(k)}\right)\right.\right.$$
$$\left.\left.+ \frac{1}{P} E\left(\vec{R}^{(k)}, N_e\right)\right]\right\}\bigg|_{\vec{R}^{(1)} = \vec{R}^{(P+1)}}$$

(11)

where the subscript "qtm" means quantum, $P$ is the number of beads, $\vec{R}^{(k)}$ denotes the particles' positions of the ring-polymer bead with index number $k$, $\Lambda_P$ is the prefactor generated by the Gaussian integral when dealing with the momentum operators in the PI formulation and $\omega_P = \sqrt{P}/\beta\hbar$ is the chain frequency of the adjacent harmonic coupling in the ring-polymer model. The above formalism describes a quantum system through an isomorphic classical ring-polymer model made of multiple beads (each bead refers to a classical system's configuration) with harmonic couplings between adjacent beads. For a finite number of $P$ beads, the formal potential of the partition function $Q_P(\beta, N_e)$ in **Eq. 11** is:

$$\phi\left(\vec{R}^{(1)}, \ldots, \vec{R}^{(P)}, N_e\right) = \sum_{k=1}^{P} \left[\frac{1}{2}\omega_P^2 \left(\vec{R}^{(k+1)} - \vec{R}^{(k)}\right)^T \underline{m}\left(\vec{R}^{(k+1)} - \vec{R}^{(k)}\right) + \right.$$
$$\left.\frac{1}{P} E\left(\vec{R}^{(k)}, N_e\right)\right]\bigg|_{\vec{R}^{(1)} = \vec{R}^{(P+1)}}$$

(12)

which is called the *effective potential*.

We define the centroid of multiple beads as $\vec{R}$, and usually the RC of a quantum system is a function with respect to the centroid of multiple beads $q = f(\vec{R})$. The coordinate transformation to decouple the centroid has the form:

$$\vec{R} = \frac{1}{P}\sum_{k=1}^{P} \vec{R}^{(k)}$$
$$\vec{R}_\Delta^{(k)} = \vec{R}^{(k+1)} - \vec{R}^{(k)}, \quad k = 1, \ldots, P-1$$

(13)

In this situation, the probability density of $q = s$ is transformed into

$$P_P(s) = \frac{\Lambda_P}{Q_P(\beta, N_e)} \int d\vec{R}^{(1)} \ldots d\vec{R}^{(P)} \exp\left[-\beta\phi\left(\vec{R}^{(1)}, \ldots, \vec{R}^{(P)}, N_e\right)\right] \delta(q - s)$$

$$= \frac{\Lambda_P}{Q_P(\beta, N_e)} \int dq d\vec{q}_{\text{trans}} d\vec{q}_{\text{primit}} d\vec{R}_\Delta^{(1)} \dots d\vec{R}_\Delta^{(P-1)} \delta(q-s) \mathcal{J}(q, \vec{q}_{\text{trans}})$$

$$\times \exp\left[-\beta \tilde{\phi}\left(q, \vec{q}_{\text{trans}}, \vec{q}_{\text{primit}}, \vec{R}_\Delta^{(1)}, \dots, \vec{R}_\Delta^{(P-1)}, N_e\right)\right]$$

$$= \frac{\Lambda_P}{Q_P(\beta, N_e)} \int d\vec{q}_{\text{trans}} \mathcal{J}(s, \vec{q}_{\text{trans}}) \int d\vec{q}_{\text{primit}} \int d\vec{R}_\Delta^{(1)} \dots d\vec{R}_\Delta^{(P-1)}$$

$$\times \exp\left[-\beta \tilde{\phi}\left(s, \vec{q}_{\text{trans}}, \vec{q}_{\text{primit}}, \vec{R}_\Delta^{(1)}, \dots, \vec{R}_\Delta^{(P-1)}, N_e\right)\right]$$

(14)

where $\left\{\vec{R}_\Delta^{(k)}\right\}_{k=1}^{P-1}$ are sampled via the staging transformation MC scheme[55] with $\vec{q}_{\text{trans}}$ and $\vec{q}_{\text{primit}}$ fixed. The evolution of $\vec{q}_{\text{trans}}$ and $\vec{q}_{\text{primit}}$ has been discussed in the above section. We further note that the force of the centroid under the *effective potential* is just the average force of all beads at their specific configurations.

**Details of the mean force estimator**

The free energy (potential of mean force) along a defined RC is a function of the probability density $P(s)$

$$F(s) = -\frac{1}{\beta} \ln P(s)$$

(15)

The free energy change between RC values $s_1, s_2$ can be computed by integrating the mean force along the RC range $[s_1, s_2]$

$$F(s_2) - F(s_1) = \int_{s_1}^{s_2} \frac{dF}{ds} ds = \int_{s_1}^{s_2} \left\langle \left(\frac{dF}{ds}\right)_{\text{estm}} \right\rangle_s^{\text{cond}} ds$$

(16)

where $\langle \cdot \rangle_s^{\text{cond}}$ represents the conditional ensemble average with the constraint $q = s$. According to the TI method, the general form of the estimator for the mean force is

$$\left(\frac{dF}{ds}\right)_{\text{estm}} = \frac{\partial \tilde{E}}{\partial q} - k_B T \frac{\partial}{\partial q} \ln \mathcal{J}(\vec{q})$$

(17)

We consider two types of RCs in this study, and the specific formula of the estimator for them can be derived easily by the chain rule.

a. For $q = |\vec{R}_R - \vec{R}_L|$, the RC is defined as the distance between two particles labeled by L and R, and $\vec{R}_L$ and $\vec{R}_R$ are their positions. In this situation, the estimator is

$$\left(\frac{dF}{ds}\right)_{estm} = \frac{(\vec{f}_L - \vec{f}_R)(\vec{R}_R - \vec{R}_L) - 4k_BT}{2|\vec{R}_R - \vec{R}_L|}$$

(18)

b. For $q = |\vec{R}_R - \vec{R}_M| - |\vec{R}_M - \vec{R}_L|$, the RC is defined as the difference of the distance between two particles labeled by L and M and the distance between two particles labeled by M and R, and $\vec{R}_L, \vec{R}_M, \vec{R}_R$ are their positions. In this situation, the estimator is

$$\left(\frac{dF}{ds}\right)_{estm} = \frac{1}{2}\left[\frac{\vec{f}_L(\vec{R}_L - \vec{R}_M) + 2k_BT}{|\vec{R}_L - \vec{R}_M|} - \frac{\vec{f}_R(\vec{R}_R - \vec{R}_M) + 2k_BT}{|\vec{R}_R - \vec{R}_M|}\right]$$

(19)

where $\vec{f}$. is used to represent the force on the corresponding particle in the above two formulas.

**Details of constrained PIHMC under GC sampling**

We apply the constrained GC-(PI)HMC method on our modeled interface system at room temperature $T = 300$ K. The number of beads used in our PI calculations is 16 when we consider the NQEs of PCET steps.[17,56] Tests for the NQEs calculations in terms of beads number convergence and temperature effect can be found in **SI Section 7**. In order to obtain more reliable statistical results, we perform 15 times independent 300,000-step sampling trajectories at different RC values for the Volmer reaction, each time with a different initial structure. For the Heyrovsky reaction, due to its better performance in the mean force convergence compared to the Volmer case, we perform 10 times independent mean force samplings for each RC case (also consisting of 300,000 (PI)HMC steps for every sampling trajectory). Specifically, for the sensitive RC range ($q_{Heyrovsky}$ from 0.85 to 0.95 Å) corresponding to the relative sharp transition of an $H_2$ molecule formation, we run each sampling trajectory for 1,000,000 steps to ensure the reliability of the ensemble average results.

**Setup details of DFT calculations**

All DFT calculations in this work are performed using a first-principles calculation software ABACUS (version 3.4.0)[46,47] with a main feature of employing numerical atomic orbitals (NAO) as the basis set, which is capable of performing efficient DFT calculations for more than hundreds of atoms in a supercell. In addition, the function of an adjustable total number of electrons with a compensating charge plate, whose position is also configurable, is implemented in the ABACUS code (**SI Section 2**). We thus employ ABACUS to perform all of the *ab-initio* calculations for labeling the DP-$N_e$ training dataset throughout the work.

The norm-conserving pseudopotentials are adopted with the valence electron configurations: [H]$1s^1$, [O]$2s^22p^4$, and [Pt]$5s^25p^65d^86s^2$. We use the generalized gradient approximation (GGA) in the form of the Perdew–Burke–Ernzerhof (PBE) version[57] for describing the exchange–correlation functional. Specifically, we choose 2s1p, 2s2p1d, and 4s2p2d1f NAO basis sets with radius cutoffs as 6, 7, and 7 Bohr respectively for H, O, and Pt elements. The kinetic energy cutoff is set to 100 Ry (1360 eV). We employ the PBC for modeling the Pt/H$_2$O interface supercell, and the k-point mesh for sampling the Brillouin zone of the slab model with dimensions of 14.06 Å × 14.06 Å × 30.89 Å is set as 2 × 2 × 1. A dipole correction[58] is included in our DFT calculations as well due to the net dipole of the interface model especially at charged states representing an electrochemical reducing condition. We use the Gaussian smearing method with a width of 0.02 Ry. We also apply the Grimme's D3 dispersion correction[59] to take account of the long-range van de Waals interaction effect.

**Training process of our DP-$N_e$ MLP**

The MLP construction is conducted by the automatic configuration-exploration workflow DP-GEN[60] in this work. Concurrent learning processes in the DP-GEN contain a series of iterations, each of which is composed of three steps: training the neural network that describes force field models, exploring the configurational space (i.e., the extended [$\vec{R}$,$N_e$] space in our work), and labeling the newly added candidate configurations by first-principles calculations, which are selected based on evaluation of the current machine learning model's accuracy. Our DP-$N_e$ model is generated according to the workflow displayed in **Figure 1(b)**. We incorporate the total number of electrons $N_e$ directly into the input layer of the neural network and use the software DeepMD-kit[61–65] to train the DP-$N_e$ MLP. All exploration trajectories are performed under the GC ensemble distribution using our GC-HMC methods. More details about the DP-GEN iterations can be found in **SI Section 3**.

## Acknowledgements

The authors gratefully acknowledge funding support from the National Natural Science Foundation of China (grant no. 52273223), Ministry of Science and Technology of the People's Republic of China (grant no. 2021YFB3800303), DP Technology Corporation (grant no. 2021110016001141), School of Materials Science and Engineering at Peking University, and the AI for Science Institute, Beijing (AISI). The computing resource of this work was provided by the Bohrium Cloud Platform (https://bohrium.dp.tech), which was supported by DP Technology.

## Competing interests

The authors declare no competing interests.

## Supporting information

Issues with the GC-DFT method specifically for the grand canonical path integral simulations. Tests for the implementation of a compensating charge scheme. DP-GEN setup parameters and tests for the accuracy of DP-$N_e$. Relationship between work function and numerical derivative $\Delta E/\Delta N_e$ of configurations in the extended $[\vec{R},N_e]$ space. Statistical distribution of $N_e^{\text{extra}}$. Total electron number change with respect to potential variation. Tests for the NQEs calculations in terms of beads number convergence and temperature effect. Miscellaneous technical details in our sampling calculations. Convergence of mean forces and potential energies with respect to HMC steps. Mean forces with respect to reaction coordinates. Tafel step results.

## Code availability

The constrained GC-PIHMC code developed in this work is available at our group's GitHub page "https://github.com/sxu39/GC-Constrained-PIHMC".

# Supplemental Information

# "A Machine-Learning Accelerated Grand Canonical Sampling Framework for Nuclear Quantum Effects in Constant Potential Electrochemistry"

Menglin Sun[1]†, Bin Jin[1]†, Xiaolong Yang[1]†, Shenzhen Xu[1,2]*

[1]School of Materials Science and Engineering, Peking University, Beijing 100871, People's Republic of China

[2]AI for Science Institute, Beijing 100084, People's Republic of China

## S1. Issues with the GC-DFT method specifically for the grand canonical path integral simulations

During the path integral (PI)[1,2] simulations under a grand canonical (GC) ensemble condition, the GC-DFT approach[3] varies the total electron number $N_e$ for each ring-polymer bead configuration to maintain the same work function, resulting in an issue of distinct $N_e$ values for different beads. This violates the requirement of the same electron number for all ring-polymer beads in the PI formalism, which is proved as follows.

Under the adiabatic approximation, we can define the state of the investigated system as $|n(\vec{R})\rangle|\vec{R}\rangle$, where $|\vec{R}\rangle$ and $|n(\vec{R})\rangle$ denote the nuclear position state and the electronic adiabatic state at nuclear position $\vec{R}$, respectively. Since we only consider the ground state for electrons' degrees of freedom (the more subtle non-adiabatic effect is beyond the scope of our current work and worth of studying in future), the adiabatic state with a variable number of electrons can be simplified into $|N_e(\vec{R})\rangle$. Given this condition, the GC ensemble quantum configuration partition function can be derived like that introduced in the main text Methods section by adding trace operation on the electronic adiabatic states:

$$\Xi_{\text{qtm}}(\beta, \mu_e) = \text{Tr}_n\{\text{Tr}_e[\exp(-\beta(\widehat{\mathcal{H}} - \mu_e \widehat{N}_e))]\}$$

$$= \lim_{P \to \infty} \text{Tr}_n\left\{\text{Tr}_e\left[\exp(-\beta(\widehat{\mathcal{H}} - \mu_e \widehat{N}_e)/P)^P\right]\right\}$$

$$
\begin{aligned}
&= \lim_{P\to\infty} \int d\vec{R}^{(1)} \dots d\vec{R}^{(P)} \sum_{N_e^{(1)}\dots N_e^{(P)}} \langle\vec{R}^{(1)}|\langle N_e^{(1)}(\vec{R}^{(1)})|\exp(-\beta(\hat{\mathcal{H}}-\mu_e\hat{N}_e)/P)|N_e^{(2)}(\vec{R}^{(2)})\rangle|\vec{R}^{(2)}\rangle \\
&\quad \times \dots \times \langle\vec{R}^{(P-1)}|\langle N_e^{(P-1)}(\vec{R}^{(P-1)})|\exp(-\beta(\hat{\mathcal{H}}-\mu_e\hat{N}_e)/P)|N_e^{(P)}(\vec{R}^{(P)})\rangle|\vec{R}^{(P)}\rangle \\
&\quad \times \langle\vec{R}^{(P)}|\langle N_e^{(P)}(\vec{R}^{(P)})|\exp(-\beta(\hat{\mathcal{H}}-\mu_e\hat{N}_e)/P)|N_e^{(1)}(\vec{R}^{(1)})\rangle|\vec{R}^{(1)}\rangle \\
&= \lim_{P\to\infty} \int d\vec{R}^{(1)} \dots d\vec{R}^{(P)} \sum_{N_e^{(1)}\dots N_e^{(P)}} \langle\vec{R}^{(1)}|\exp(-\beta\hat{\mathcal{H}}_{N_e^{(2)}}/P)|\vec{R}^{(2)}\rangle \\
&\quad \times \dots \times \langle\vec{R}^{(P-1)}|\exp(-\beta\hat{\mathcal{H}}_{N_e^{(P)}}/P)|\vec{R}^{(P)}\rangle \langle\vec{R}^{(P)}|\exp(-\beta\hat{\mathcal{H}}_{N_e^{(1)}}/P)|\vec{R}^{(1)}\rangle \\
&\quad \times \exp\left(\beta\mu_e/P \sum_{k=1}^{P} N_e^{(k)}\right) \langle N_e^{(1)}(\vec{R}^{(1)})|N_e^{(2)}(\vec{R}^{(2)})\rangle \\
&\quad \times \dots \times \langle N_e^{(P-1)}(\vec{R}^{(P-1)})|N_e^{(P)}(\vec{R}^{(P)})\rangle \langle N_e^{(P)}(\vec{R}^{(P)})|N_e^{(1)}(\vec{R}^{(1)})\rangle
\end{aligned}
$$

(S1)

The electronic ground states with different numbers of electrons are orthogonal in the Fock space,[4] the inner product therefore satisfies the equation when $P \to \infty$:

$$\langle N_e^{(i)}(\vec{R}^{(i)})|N_e^{(j)}(\vec{R}^{(j)})\rangle = \delta_{N_e^{(i)},N_e^{(j)}}$$

(S2)

Based on the above property, the numbers of electrons for all beads configurations should be the same in the PI formalism and the GC ensemble partition function can be expressed with the canonical ensemble partition function as:

$$
\begin{aligned}
\Xi_{\text{qtm}}(\beta,\mu_e) &= \sum_{N_e} \exp(\beta\mu_e N_e) \lim_{P\to\infty} \int d\vec{R}^{(1)} \dots d\vec{R}^{(P)} \langle\vec{R}^{(1)}|\exp(-\beta\hat{\mathcal{H}}/P)|\vec{R}^{(2)}\rangle \\
&\quad \times \dots \times \langle\vec{R}^{(P-1)}|\exp(-\beta\hat{\mathcal{H}}/P)|\vec{R}^{(P)}\rangle \langle\vec{R}^{(P)}|\exp(-\beta\hat{\mathcal{H}}/P)|\vec{R}^{(1)}\rangle \\
&= \sum_{N_e} \exp(\beta\mu_e N_e) Q_{\text{qtm}}(\beta, N_e)
\end{aligned}
$$

(S3)

The above quantum partition function exhibits the same form as the classical one presented in the main text **Eq. 1**. Here we want to stress again that for a GC ensemble sampling, no matter in a classical or a quantum situation, the instantaneous work function or chemical potential of each sampled configuration is never a proper controlling parameter or a degree of freedom incorporated into the potential energy surface (PES), on the other hand, its conjugate thermodynamic variable – the particle number (i.e., $N_e$ in our study) – is the appropriate choice. This basic argument is supported by the fundamental principle of statistic mechanics, regardless of the debate that how we can better describe electric double layers at interfaces, using surface charge density or work function.

**S2. Tests for the implementation of a compensating charge scheme**

An important prerequisite for the realization of GC sampling to achieve a constant potential condition is that, we should be able to perform electronic structure calculations of the interface system with variable number of electrons $N_e$. The functions of implementing a charge compensation scheme[5] (keeping neutrality of the periodic supercell) and a dipole correction method[6] are realized in the first-principles calculation software ABACUS (Atomic-orbital Based Ab-initio Computation at UStc),[7,8] so that we could add/extract electrons into/from the interface model with a reasonable position of a compensating charge plate to mimic an electric double layer at the Pt/H$_2$O interface. We place a compensating charge plate in the vacuum region (~ 2 Å above the water solvation layer), and a dipole correction is included at 7.8 Å above the compensating charge plate (in the vacuum) in the interface model for our DFT calculations (illustrated in **Figure 2(a)** in the main text).

We output the electrostatic energy of an electron along the direction (z) perpendicular to the Pt surface, for a testing system with added electron number $N_e^{\text{extra}} = 0.2$. The testing system is a (3×3) Pt (111) surface slab composed of three atomic layers illustrated in **Figure S1**. The modeled electrode surface contains 27 Pt atoms with 1 monolayer (ML) hydrogen coverage and includes 6 explicit water molecules. We place a compensating charge plate in the vacuum region above the water layer and a dipole correction is included in DFT calculations as well. We can see from **Figure S1** that the electrostatic energy results from ABACUS are in perfect agreement with those produced by the well-established *ab-initio* code Quantum ESPRESSO (QE)[9,10], which justifies our implementation of an effective electric double layer near the electrode surface adapted for adjustable total number of electrons in our atomic interface model.

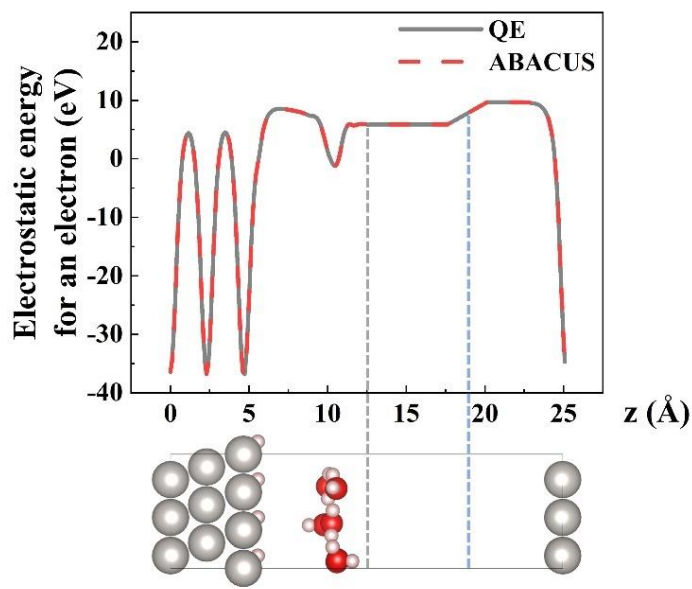

**Figure S1**. Electrostatic energy of an electron along the direction (z) perpendicular to the Pt surface calculated by QE (Quantum ESPRESSO) and ABACUS. The gray vertical dashed line represents the position of the compensating charge plate and the blue dashed line corresponds to the dipole correction. A schematic structure plot of the testing interface model is shown aligned below the average electrostatic energy curves. Gray, red, and white spheres correspond to platinum, oxygen, and hydrogen atoms, respectively.

## S3. DP-GEN setup parameters and tests for the accuracy of DP-$N_e$

The concurrent learning process in the DP-GEN workflow[11] goes through a series of iterations consisting of training, exploration, and labeling steps as introduced in the main text **Methods** section. In the training stage, four models with different random seeds are trained using DeePMD-kit[12,13] with $4\times10^5$ steps. The embedding network has three layers with 25, 50 and 100 nodes and the fitting network is composed of three layers, each of which has 240 nodes. The loss functions are minimized with an exponentially decay learning rate from $1.00 \times 10^{-3}$ to $3.51\times10^{-8}$. In the exploration stage, when the model deviation of one configuration is in the range [0.10, 0.60] eV/Å, this structure will be identified as a candidate structure, further labeled by *ab-initio* calculations, and added to the dataset in the next iteration. If model deviations of more than 85% of all snapshots in a $10^4$-step trajectory are lower than 0.10 eV/Å, the DP-GEN iterations can be regarded as converged.

We use the converged DP-$N_e$ machine learning potential (MLP) to run sampling trajectories at different reaction coordinates (RC) under the reduction voltage $U = -0.9$ V vs SHE in classical situations and every single sampling trajectory consists of 300,000 steps. Then 250 snapshots for the Volmer reaction and 150 snapshots for the Heyrovsky reaction are randomly selected as the testing dataset. The predicted energies and forces from the DP-$N_e$ potential match well with those from DFT calculations on the testing dataset as shown in **Figure S2**.

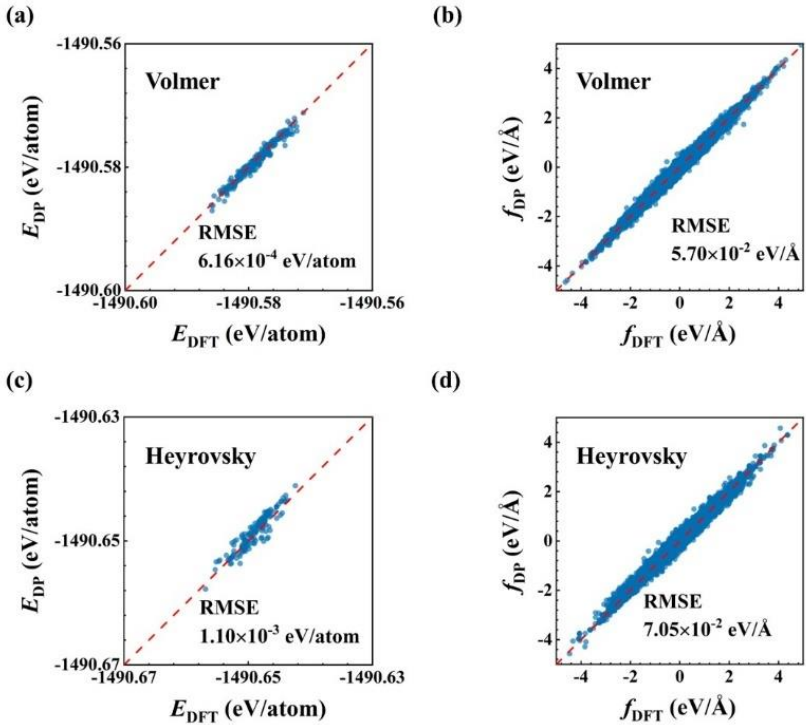

**Figure S2**. Comparisons of energies ($E_{DP}$ vs $E_{DFT}$) and forces ($f_{DP}$ vs $f_{DFT}$) obtained by the DP-$N_e$ and DFT on the testing dataset. Panels (a) and (b) correspond to the Volmer step, and panels (c) and (d) show the results of the Heyrovsky step. The total root mean square errors (RMSE) of energies and forces are listed in the corresponding panels.

## S4. Relationship between work function and numerical derivative $\Delta E/\Delta N_e$ of configurations in the extended $[\vec{R}, N_e]$ space

In this section, we employ DFT calculations to demonstrate the equivalence of work function $W(\vec{R}, N_e)$ to numerical derivative $-\frac{\Delta E(\vec{R}, N_e)}{\Delta N_e}$ for several configurations with different electron numbers (**Figure S3(a)**). The work function $W(\vec{R},N_\text{e})$ here is calculated by $W = E_\text{vac} - E_\text{Fermi}$, where $E_\text{vac}$ is the electronic energy in vacuum, and $E_\text{Fermi}$ is the Fermi level of the modeled metal system. This fundamental relationship lays the foundation for our grand canonical hybrid Monte Carlo (GC-HMC) method to efficiently conduct the ensemble average of the interface model's work function.

We conduct tests on six distinct systems, five of which are (5×5) Pt (111) surface slab composed of four atomic layers (with $N_\text{e}^\text{extra} = 0, 0.5, 1, 1.5$, and 2, respectively), and one is a (3×3) Pt (111) slab modeled with three Pt layers with $N_\text{e}^\text{extra} = 0$. The spacing electron number ($N_\text{e}^\text{s}$) for finite differentiation is set to 0.001. Then we evaluate the energy numerical derivative with respect to $N_\text{e}$ as (using DFT calculations):

$$-\frac{\Delta E(\vec{R}, N_\text{e})}{\Delta N_\text{e}} = -\frac{E(\vec{R}, N_\text{e} + N_\text{e}^\text{s}) - E(\vec{R}, N_\text{e} - N_\text{e}^\text{s})}{2N_\text{e}^\text{s}}$$

(S4)

The results from **Figure S3(a)** clearly show that the work function of a specific configuration $[\vec{R}, N_\text{e}]$ can be derived from the derivative of $E(\vec{R}, N_\text{e})$ with respect to $N_\text{e}$.

Moreover, we conduct testing calculations to examine the relationship between the analytic derivative $\frac{\partial E(\vec{R},N_\text{e})}{\partial N_\text{e}}$ and the numerical derivative $\frac{\Delta E(\vec{R},N_\text{e})}{\Delta N_\text{e}}$ of the DP-$N_\text{e}$ energy for sampled configurations, where $\frac{\partial E(\vec{R},N_\text{e})}{\partial N_\text{e}}$ is an additional output obtained from the automatic differentiation implemented in our DP-$N_\text{e}$ MLP construction. We randomly select six configurations of a (5×5) Pt slab sampled in the extended $[\vec{R},N_\text{e}]$ space with $N_\text{e}^\text{extra} = 0.43, 0.53, 0.63, 0.73, 0.77$ and $0.78$, respectively. The energy numerical derivative $\frac{\Delta E(\vec{R},N_\text{e})}{\Delta N_\text{e}}$ is also calculated according to **Eq. S4** with $N_\text{e}^\text{s} = 0.001$, and the energy $E(\vec{R},N_\text{e})$ here corresponds to the inferred energy from DP-$N_\text{e}$. We find that $\frac{\partial E(\vec{R},N_\text{e})}{\partial N_\text{e}}$ and $\frac{\Delta E(\vec{R},N_\text{e})}{\Delta N_\text{e}}$ are in perfect agreement as shown in **Figure S3(b)**. We thus can simply use the energy derivative of each sampled microstate with respect to $N_\text{e}$ as the estimator to do the ensemble average calculations of the interface model's work function, which can be easily obtained from the automatic differentiation implemented in MLP construction.

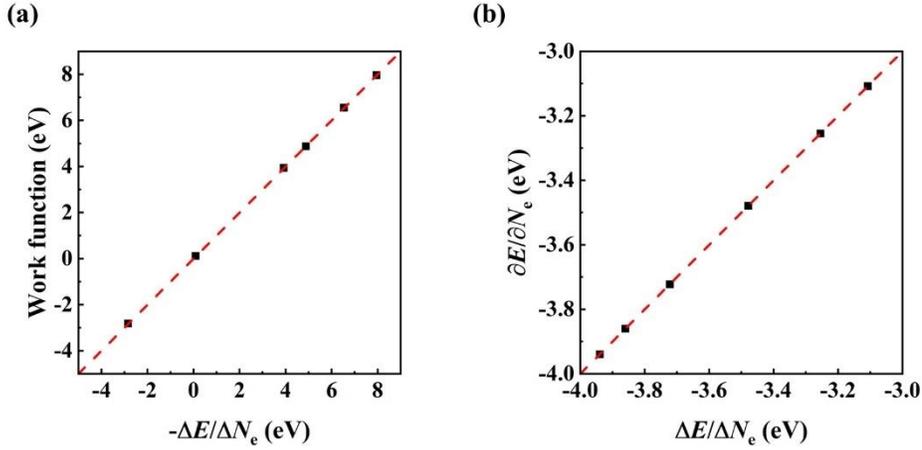

**Figure S3**. (a) Comparison between work function and numerical derivative $-\frac{\Delta E(\vec{R},N_e)}{\Delta N_e}$ of the testing configurations in the $[\vec{R},N_e]$ space by DFT calculations. (b) Comparison between analytic derivative $\frac{\partial E(\vec{R},N_e)}{\partial N_e}$ and numerical derivative $\frac{\Delta E(\vec{R},N_e)}{\Delta N_e}$ of the testing configurations in the $[\vec{R},N_e]$ space from our DP-$N_e$ MLP inference.

## S5. Statistical distribution of $N_e^{extra}$

We plot a histogram to present the statistical distribution of the interface system's electron number (in the form of $N_e^{extra}$ added into our interface model) at $\mu_e$ = -3.5 eV (U = -0.9 V vs SHE) and RC $q_{Volmer}$ = -0.21 Å, for the Volmer PCET step in a classical situation. The results exhibit a normal distribution, indicating a reasonable electron number fluctuation of the interface system in GC sampling simulations.

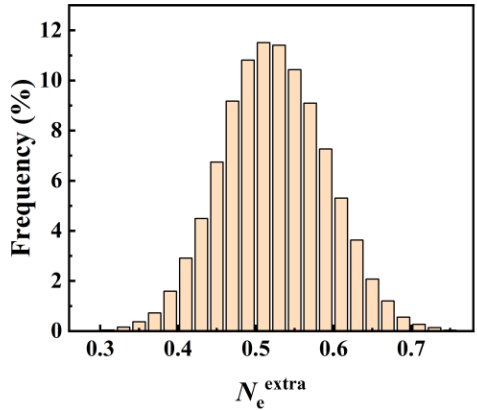

**Figure S4**. Statistical distribution of the modeled interface system's electron number (represented by $N_e^{extra}$) at $U$ = -0.9 V vs SHE, RC $q_{Volmer}$ = -0.21 Å for the Volmer step under a classical situation.

## S6. Total electron number change with respect to potential variation

We show the dependence of our interface model's $N_e$ on applied potentials in this section, particularly focusing on the initial state (IS) and final state (FS) of the Volmer step as shown in **Figure S5**. Please note that we use the extra electron number $N_e^{extra}$ added/subtracted into/from the interface model to represent $N_e$, in a similar way as we do in the main text **Figure 3(c)**. When potentials become more negative, an increase in the total electron number is observed due to the electrochemical driving force requiring more electrons to facilitate reduction reactions. The error bars in **Figure S5** denote standard errors of average $N_e^{extra}$ derived from 15 times independent simulations for each case at a specific applied potential condition.

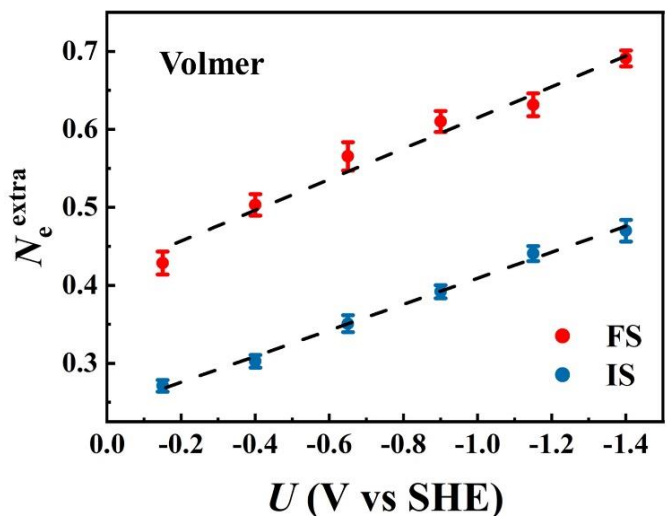

**Figure S5**. Dependence of our interface model's total electron number (represented by $N_e^{extra}$) on applied potentials, specifically for the PCET Volmer step's IS and FS under a classical situation. Error bars on each data point indicate standard errors from 15 times independent simulations.

## S7. Tests for the NQEs calculations in terms of beads number convergence and temperature effect

The total beads number ($P$) in PI simulations is typically set as 16 at room temperature ($T$ = 300 K) based on earlier studies,[14,15] we further perform beads number convergence tests in this section to validate our PI calculations. We increase $P$ from 16 to 32 in the GC-PIHMC simulation for the Volmer step at $U = -0.9$ V vs SHE and make comparison of the free energy profile. The good quantitative agreement between $P = 16$ and $P = 32$ cases shown in **Figure S6(a)** justifies the sufficiency of setting the beads number as 16 in our PI calculations at room temperature.

Since NQEs is relatively sensitive to temperature change, lowering the temperature in PI simulations is expected to enhance the NQEs' impact on reaction's activation energies. We thus perform tests to examine the free energy profiles of the Volmer step at $U = -0.9$ V vs SHE under both the classical and quantum situations at different temperatures (250 K and 300 K). The results shown in **Figure S6(b) and (c)** are consistent with our expectation that as temperature $T$ decreases, the activation energy reduction induced by NQEs becomes more significant, indicating our GC-PIHMC method exhibits a qualitatively correct behavior in NQEs simulations.

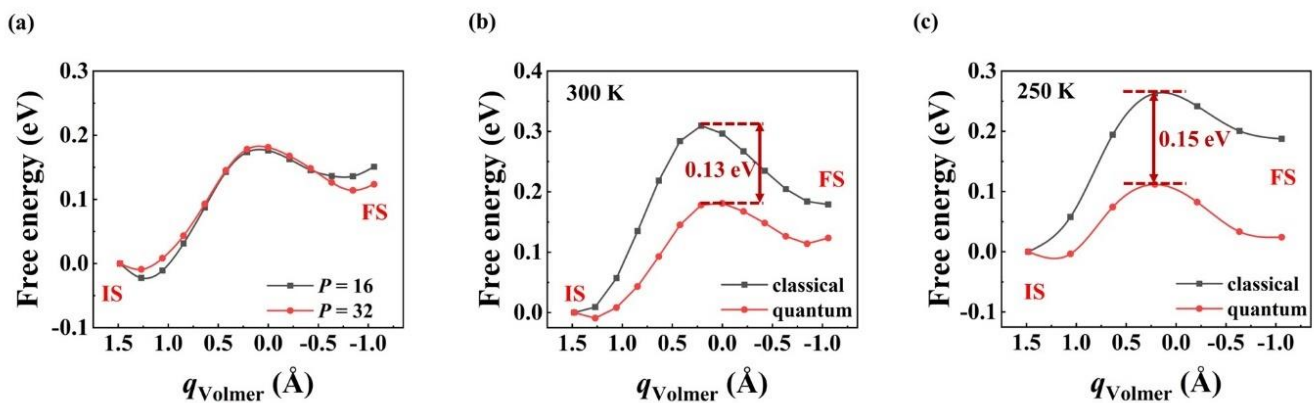

**Figure S6**. (a) Free energy profiles of the Volmer step with beads number $P = 16$ and 32 in our GC-PIHMC simulation at U = -0.9 V vs SHE. (b) and (c) Free energy profiles of the Volmer step under both the classical and quantum situations at different temperatures (300 K and 250K) at $U = -0.9$ V vs SHE.

### S8. Miscellaneous technical details in our sampling calculations

We implement several constraints on atomic coordinates to ensure reasonable configurational sampling in our GC-(PI)HMC simulations. We place a rigid wall ~1.1 Å below the compensating charge plate to prevent water molecules from diffusing to the vacuum region. As we only allow the transferring proton (originally in $H^+_{sol}$ form) to move from IS to FS along our constrained HMC samplings in PCET free energy calculations, we have to make sure that H* atoms (except for the H* involved in the Heyrovsky step) at the Pt surface would not migrate to water layers. We thus introduce another rigid barrier, specifically for the adsorbed H* atoms, to prevent them from going upwards into the explicit $H_2O$ molecule region. In addition, we place one more rigid wall, only for the solvation $H_2O$ molecules, to avoid water adsorption at Pt sites which may disturb our free energy calculations of PCET reactions.

In our GC-PIHMC sampling simulations, trial moves for different types of degrees of freedom are selected based on a preset ratio. We set the probabilities of making trial moves for the internal degrees of freedom within the quantized beads' configurations in PIMC ($\vec{R}_{\Delta}^{(k)}$), centroid atomic coordinates ($\vec{R}$) and total number of electrons ($N_e$) as 0.32, 0.28, and 0.40, which corresponds to $a_1 = 0.32$ and $a_2 = 0.60$ mentioned in **Figure 1(a)** in the main text.

**S9. Convergence of mean forces and potential energies with respect to HMC steps**

We present in **Figure S7** the convergence of potential energies and mean forces with respect to the HMC steps for different cases of the Volmer and Heyrovsky paths in both of classical and quantum situations. All results are obtained at a specific reduction potential with $U$ = -0.9 V vs SHE.

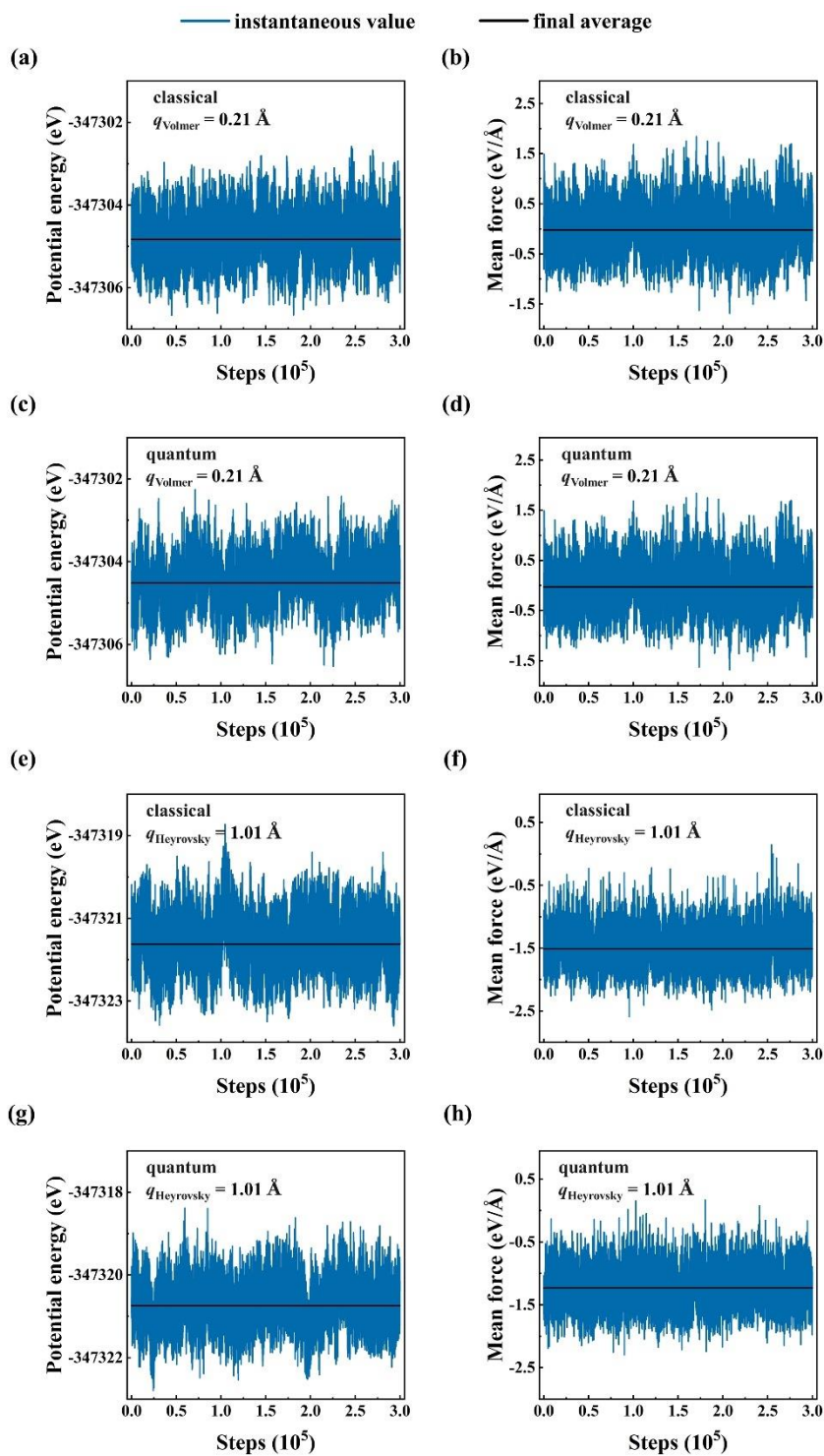

**Figure S7.** (a)(c)(e)(g) Potential energies and (b)(d)(f)(h) mean forces fluctuation with respect to (PI)HMC steps in both of classical and quantum situations at $q_{\text{Volmer}} = 0.21$ Å for the Volmer reaction and $q_{\text{Heyrovsky}} = 1.01$ Å for the Heyrovsky reaction. All results are obtained at a reduction potential $U = -0.9$ V vs SHE. Blue curves show the instantaneous values, and black lines are the final averages along complete sampling trajectories.

## S10. Mean forces with respect to reaction coordinates

We display in **Figure S8** the mean force values with respect to RC for the Volmer and Heyrovsky reaction at $\mu_e$ = -3.5 eV ($U$ = -0.9 V vs SHE) in both of classical and quantum situations. Error bars on each data point refer to standard errors from 15 (for Volmer path) and 10 (for Heyrovsky path) independent simulations for each RC condition.

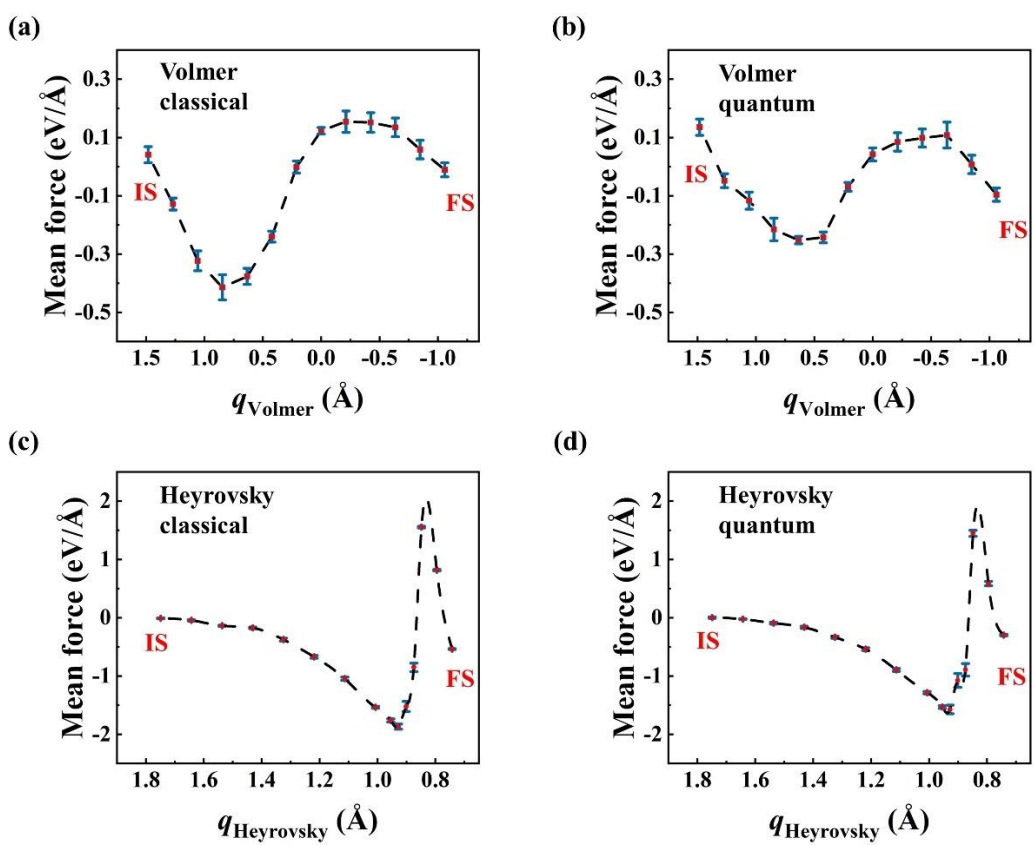

**Figure S8**. Mean forces change along the reaction path from IS to FS for the (a)(b) Volmer and (c)(d) Heyrovsky reaction at $U$ = -0.9 V vs SHE in both the classical and quantum situations. Error bars represent standard errors from 15 (Volmer) or 10 (Heyrovsky) independent simulations. Black dashed curves are simply guidelines showing the trend of mean force variation with respect to RC.

## S11. Tafel step results

We investigate the Tafel step (H* + H* → $H_2$), which is an elementary step involved in the HER characterized as a chemical (instead of an electrochemical) reaction with negligible charge transfer across the interface. The non-electrochemical Tafel step is expected to exhibit negligible dependence on applied potentials, we thus perform samplings under a canonical ensemble with the constrained (PI)HMC method on our modeled interface system with a traditional DP MLP model (instead of our developed DP-$N_e$ MLP introduced before).

We employ ABACUS[7,8] to perform all of the *ab-initio* calculations. The Perdew–Burke–Ernzerhof (PBE) exchange correlation functional[16] with a 3 × 3 × 1 k-point mesh and a 100 Ry (1360 eV) kinetic energy cutoff is used for all single-point calculations. We construct a (4×4) supercell model of the Pt(111) surface composed of four atomic layers with 1 ML hydrogen coverage at Pt atop sites as illustrated in **Figure S9(a)**. Pt atoms of the bottom layer is fixed during DFT calculations and sampling processes. A vacuum region of 15 Å thickness is added above the adsorbed hydrogen atoms in order to decouple periodic images of the slab model. We test the implicit solvation effect on the energy barrier of the Tafel step using the climbing image nudged elastic band (CI-NEB) method.[17] The potential energy barrier is 0.71 eV when the implicit solvation effect is taken into account, and 0.64 eV without the implicit solvation, indicating a small difference of 0.07 eV. We therefore do not include the implicit solvation model in our simulations due to the extra computational cost. This setup is also consistent with a previous theoretical work.[18,19]

For the Tafel reaction, we define the RC $q_\text{Tafel}$ as the distance $|\vec{r}_\text{HH}|$ between the two hydrogen atoms forming the H$_2$ molecule (**Eq. S5**), this definition is schematically depicted in **Figure S9(b)**.

Tafel RC: $$q_\text{Tafel} = |\vec{r}_\text{HH}| \qquad (S5)$$

When considering NQEs, the above two hydrogen atoms are quantized by ring-polymer beads and $q_\text{Tafel}$ is defined by the distance of the beads' centroids (illustrated in the right panel of **Figure S9(b)**). The number of beads used in our PI calculations is also 16 at room temperature $T = 300$ K.

Our DP model is trained based on a 1253-structure dataset generated by DP-GEN.[11] In the exploration stage of the DP training process, if the model deviation of one configuration is in the range [0.10, 0.30] eV/Å, this structure will be identified as a candidate. If model deviations of more than 95% of all snapshots in a $10^4$-step trajectory are lower than 0.10 eV/Å, the DP-GEN iterations can be regarded as converged. The accuracy of the DP model is tested on a 270-structure testing dataset which contains a series of configurations with different RCs along the reaction path. The predicted energies and forces from our DP model match well with those from DFT calculations as shown in **Figure S9(c)**. We perform independent samplings for each RC case consisting of 1,000,000 (PI)HMC steps for every sampling trajectory. We present the fluctuations of potential energies and mean forces with respect to the (PI)HMC steps for the Tafel path in both of classical and quantum situations in **Figure S10**. **Figure S11** shows the calculated free energy profiles with or without NQEs. The activation energy of the Tafel step is 0.536 eV and 0.537 eV for classical and quantum cases, respectively. The classical activation energy result is in good agreement with a recent computational work[20], where a similar activation energy of the Tafel step (0.53 eV) was reported. We can see clearly in **Figure S11** that the quantitative influence on the Tafel step's free energy profile is negligible when incorporating NQEs, indicating that the tunneling mechanism may not facilitate the Tafel reaction step.

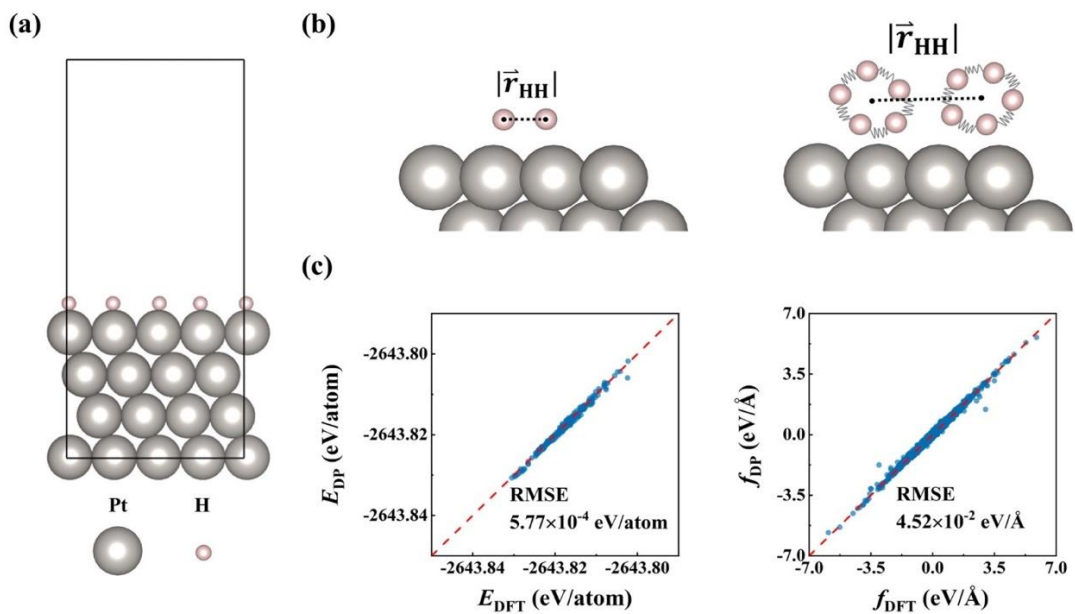

**Figure S9.** (a) Side view of the simulated (4×4) Pt (111) surface slab composed of four atomic layers containing a 1 ML adsorbed hydrogen. (b) Illustration of the key atoms relevant to the RC definitions for classical (left) and quantum (right) situations. Adsorbed hydrogen atoms are not displayed in these schematic plots for clarity. (c) Comparisons of the energies ($E_{DP}$ vs $E_{DFT}$) and forces ($f_{DP}$ vs $f_{DFT}$) obtained by the DP and DFT methods on the testing dataset. The total root mean square errors (RMSE) of energies and forces are listed in the corresponding panels.

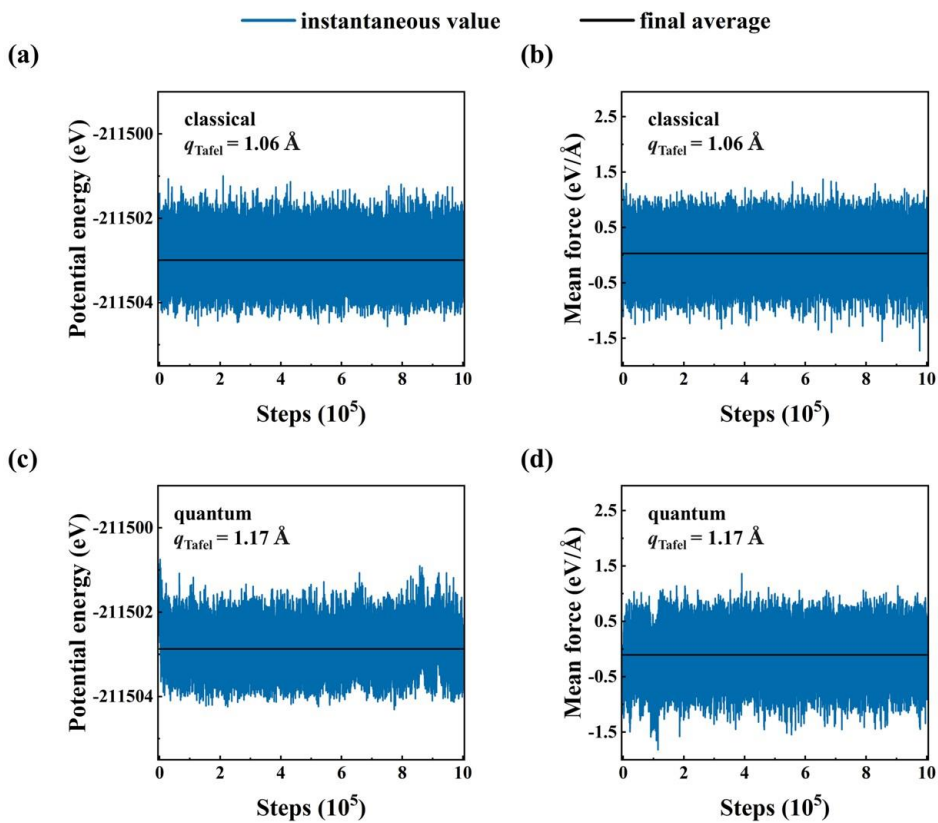

**Figure S10**. Potential energy and mean force fluctuation with respect to (PI)HMC steps for the Tafel step at $q_{\text{Tafel}} = 1.06$ Å (classical situation) and $q_{\text{Tafel}} = 1.17$ Å (quantum situation). Blue curves show the instantaneous values, and black lines are the final averages along complete sampling trajectories.

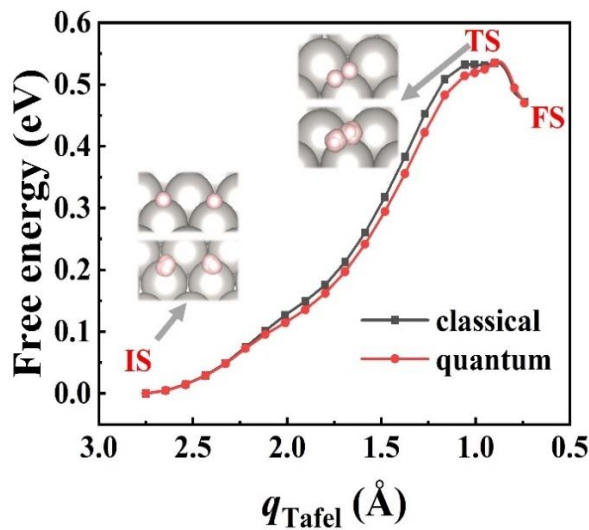

**Figure S11**. Classical and quantum free energy profiles with respect to the defined RC of the Tafel step. The schematic structural plots represent the IS and TS for classical (upper part) and quantum (lower part) cases.